\documentclass[a4paper,twoside]{article}
\newcounter{the_style}
\input{main.macros}
\title{Sensitivity And Out--Of--Sample Error in Continuous Time Data Assimilation}
\author{Jochen Br\"{o}cker and Ivan~G. Szendro\\
Max--Planck--Institut f\"{u}r Physik komplexer Systeme\\
N\"{o}thnitzer Strasse~34\\
01187 Dresden \\
Germany}
\begin{document}
\maketitle
\begin{abstract}
Data assimilation refers to the problem of finding trajectories of a prescribed dynamical model in such a way that the output of the model (usually some function of the model states) follows a given time series of observations.
Typically though, these two requirements cannot both be met at the same time---tracking the observations is not possible without the trajectory deviating from the proposed model equations, while adherence to the model requires deviations from the observations.
Thus, data assimilation faces a trade--off.
In this contribution, the sensitivity of the data assimilation
with respect to perturbations in the observations is identified as
the parameter which controls the trade--off.
A relation between the sensitivity and the out--of--sample~error is established which allows to calculate the latter under operational conditions.
A minimum out--of--sample~error is proposed as a criterion to set an appropriate sensitivity and to settle the discussed trade--off.
Two approaches to data assimilation are considered, namely variational data assimilation and Newtonian nudging, aka synchronisation.
Numerical examples demonstrate the feasibility of the approach.
\end{abstract}

\section{Introduction}
\label{sec:introduction}
Data Assimilation is one of several names for a problem (or class of problems) which in broad terms might be described thus:
Suppose we are given a time series of observations, generated by some dynamical process.
Further, we are given a dynamical model of the considered process, derived either from first principles, data analysis or other procedure.
The task is to find orbits of the considered model which are consistent with the given time series of observations.
We will restrict ourselves here to the calculation of individual orbits, but ideally, some form of distribution of orbits given the observations would be desired.
Usually, the model evolves in some state space which is not
identical to the space in which the observations live.
Rather, it is assumed that there are unobserved degrees of
freedom, and the trajectories are supposed to follow the
observations only after some function has been applied.
Clearly, data assimilation appears in many branches of science and
engineering in various different context and guises and in
connection with different objectives.
Reasons as to why data assimilation is performed might be assessment of the assimilating model, investigation of the underlying dynamics, and to obtain initial conditions for dynamical forecasts of future observations.
Naturally, a large variety of different solutions to this problem have been proposed, which often differ in a number of important details.
There are nonetheless issues which pertain to data assimilation in general and which are independent of the specifics of a particular algorithm.
The fact that different communities have different names for data
assimilation often obscures the similarities between the
approaches.
A fundamental issue is that in all but the most fortunate situations, data assimilation will have to find a trade--off between two basic yet mutually incompatible objectives: Finding a trajectory which is close to the observations versus finding a trajectory which is close to being an orbit of the model.
The commonplace that all models are wrong implies that indeed these two objectives cannot be reached at the same time and that the trade--off is a nontrivial one.
In the present contribution, this trade--off of data assimilation will simply be referred to as {\em the trade--off}.
In this paper, a general criterion is proposed and investigated by which this trade--off can be settled.
Any such criterion obviously needs some justification, but typically a mathematical justification cannot be given without introducing additional principles or dicta which, by themselves, have no foundation other than appearing reasonable or providing useful results.
Ideally, data assimilation should provide solutions which are close to the trajectories of some imaginary system which is thought of as having generated the observation.
Unfortunately, this is not an operationally feasible approach, since such trajectories are not available, and if they were, there would be no need for data assimilation.
Even then, the trajectories of the underlying true system (if it existed) and those of the model might not even be comparable.
For instance, the respective state space dimensions of model and true system might be completely different, rendering a comparison impossible.
Instead, a minimum out--of--sample error is proposed as a criterion to settle the aforementioned trade--off in data assimilation.
This quantity is not unsimilar to the out--of--sample error concept from statistics.
The observations are assumed to be corrupted by random (although not necessarily white) perturbations.
If noise corrupted data is assimilated into the model, the result should be close to the observations, even if the latter were corrupted with a {\em different} realisation of the noise; in other words, the result should be close to hypothetical observations with independent errors.
The out--of--sample error quantifies the extent to which this is the case.
Alternatively, the out--of--sample error can be considered as the error with respect to the true (unperturbed) observations, plus a constant (the variance of the observational noise).
What renders the out--of--sample error interesting from an application point of view is that it can be estimated from quantities which are, at least in principle, available in operational situations.
More specifically, it is shown that the out--of--sample error is the sum of the tracking error and a quantity involving the so--called sensitivity.
Sensitivity concepts have already been studied in various publications on data assimilation, for example~\citet{cardinali04,liu_estimating_2008,liu_analysis_2009}, for the special case of 3D--Var and 4D--Var in discrete time.
In the present work, the concepts of sensitivity and out--of--sample error will be considered in the context of data assimilation schemes in continuous time.
The sensitivity is a generalisation of the trace of the hat matrix known from linear regression~\citep[see e.g.][]{hastie01}.
In that context, the relation between sensitivity and out--of--sample error is well known as the $C_p$--criterion.
The ideas developed in the present paper, outlined in Section~\ref{sec:oos-in-data-assimilation}, can be considered as generalisations of these concepts.
How the presented approach could be employed is explained in detail in the context of two approaches to data assimilation.
In Section~\ref{sec:variational-data-assimilation}, a general variational data assimilation technique~\citep[cf e.g.][]{ledimet86,courtier87,derber89,apte08,judd08-2,broecker_variational_2010} is considered.
This variational approach leads to two--point boundary value problems, which can be solved either sequentially in time or using the so--called collocation method.
It is demonstrated how to compute the sensitivity in both cases with minimum additional costs by reusing  quantities which have been calculated already during the data assimilation procedure proper.
As a further example, data assimilation through synchronisation \citep[aka nudging, see][]{seaman90} is
investigated in Section~\ref{sec:theory-synchronisation}.
In this situation, the sensitivity is essentially given by the coupling parameter.
Finally, numerical examples are studied in Section~\ref{sec:numerical-experiments}, demonstrating the feasibility of the approach.
\section{The out--of--sample performance of data assimilation}
\label{sec:oos-in-data-assimilation}
\subsection{A prelude}
\label{subsec:prelude}
The purpose of this subsection is to provide a smooth entry to those readers who are used to think in the context of three and four dimensional variational assimilation in discrete time. 
It might also serve as a reminder of a few basic facts from the theory of linear estimation which will be of use in later discussions.
Some notation will be introduced on the way.
Suppose we are making an observation $\eta$, which we assume to be some quantity of interest $\zeta$ plus some noise, that is, $\eta = \zeta + r$, where $r$, the noise, is random with mean $\Ex r = 0$ and variance $\Ex r^2 = s$.
An {\em estimator} is a function $y(\eta)$, with the hope that $y(\eta) \cong \zeta$.
Obviously, this cannot be exactly true.
A useful measure of estimator performance is the mean square error.
The mean square error admits an instructive decomposition (the term on the left hand side is our definition of the mean square error):
\beq{equ:p10}
\E{\zeta - y}^2 = (\zeta - \bar{y})^2 + \E{\bar{y} - y}^2
\eeq
with $\bar{y} = \Ex y(\eta)$. 
The first and second terms in~\eqref{equ:p10} are called bias and variance of $y$, respectively.
If $\bar{y} = \zeta$, the estimator is called unbiased.
An estimator is linear if $y(\eta) = H \eta$.
(Here, $H$ is but a scalar; in general, it is obviously a matrix.)
For linear estimators, $\bar{y} = H \zeta$; the variance of $y$ is then $\E{\bar{y} - y}^2 = H^2 s$.
A popular linear estimator is the least squares estimator, which obtains as the minimum of the function $A(y) = (\eta - y)^2$.
In the present trivial case, this is $y(\eta) = \eta$.
Clearly, this estimator is linear, unbiased, and has variance $s$.
The Gauss--Markov theorem states that the least squares estimator has minimum variance among all unbiased linear estimators, a property often referred to as BLUE (best linear unbiased estimator).
It is possible though to find estimators which are biased but nonetheless sometimes have a smaller mean square error than BLUE.
We said ``sometimes'' here because for a biased estimator, the mean square error depends also on the unknown $\zeta$, and so does its performance compared to BLUE.
The potential of biased estimators may be illustrated with the following example.
Suppose we are not completely clueless as to the possible values of $\zeta$, but we have a rough first guess $\eta_0$.
To employ this first guess, we determine our estimator as the minimum of the modified function
\beq{equ:p20}
A(y) = \alpha \cdot (\eta - y)^2 + (1 - \alpha) \cdot (y - \eta_0)^2,
\eeq
where $\alpha$ is a new parameter between zero and one.
Our estimator becomes 
\beq{equ:p25}
y(\eta) = \alpha \eta + (1 - \alpha) \eta_0.
\eeq
Clearly, this estimator is biased.
Roughly speaking, $\alpha$ ought to represent our confidence in the quality of the observation $\eta$ versus our confidence in the first guess $\eta_0$; for $\alpha \to 1$ (perfect confidence), we get $y(\eta) = \eta$, while $\alpha \to 0$ (no confidence) gives $y(\eta) = \eta_0$.
A simple calculation gives
\beqn{equ:p30}
\text{Mean Square Error} = (1 - \alpha)^2 (\zeta - \eta_0)^2 + \alpha^2 s.
\eeq
It is easily seen that choosing 
\beq{equ:p35}
\alpha = \frac{(\zeta - \eta_0)^2}{(\zeta - \eta_0)^2 + s}
\eeq
yields a minimum error of
\[
\frac{1}{\frac{1}{s} + \frac{1}{(\zeta - \eta_0)^2}} < s.
\]
Although the optimal $\alpha$ cannot be determined practically, since it involves the unknown quantity $\zeta$, we see that there are potentially better estimators than BLUE.
It thus seems reasonable to look for ways to determine the optimal $\alpha$ at least approximately.
We will do this presently, but first note that our estimator could alternatively be interpreted as another instance of BLUE, namely, if $\eta_0$ is regarded as a second observation of $\zeta$.
More precisely, suppose that $\eta_0 = \zeta + r_0$, with $\Ex r_0 = 0$ and $\Ex r_0^2 = s_0$.
This interpretation is often imposed in the data assimilation literature, where $\eta_0$ is termed the {\em background}.
It can be shown that the estimator in Equation~\eqref{equ:p25} is the BLUE estimator of $\zeta$, {\em provided that} $\frac{\alpha}{1 - \alpha} = \frac{s_0}{s}$.
Indeed, the estimator has Mean Square Error~$ = (1 - \alpha)^2 s_0 + \alpha^2 s$,
and this expression is minimal for the said $\alpha$.
In other words, the estimator is BLUE only if in~\eqref{equ:p20}, the two terms are weighted according to the respective error covariances of $\eta$ and $\eta_0$, which are often referred to as the {\em observation} and {\em background} error covariances, respectively.
We get from this discussion that there are different ways to motivate the minimizer of the function in Equation~\eqref{equ:p20} as an estimator; either as an estimator which is biased towards a first guess, or alternatively as an unbiased estimator where the first guess is treated as another observation.
We will again encounter this situation in the case of variational data assimilation.
Whatever the motivation for the estimator, there is the problem of having to determine a good $\alpha$, ideally one that minimizes the mean square error.
One possible approach would be to use estimators of $s$ and $s_0$ in order to set $\alpha$.
Besides being a formidable problem on its own, it would force us to interprete $\eta_0$ as a random variable.
An alternative route, pursued here, is to find a reasonable substitute for the mean square error, and subsequently approximate the optimal $\alpha$ by minimizing that substitute.
We will employ the {\em out--of--sample error} as a substitute.
Imagine we had another observation $\eta' = \zeta + r'$, with $\zeta$ still the same but with $r'$ being another realisation of the observational noise, that is $\Ex r' = 0$, $\Ex r'^2 = s$, but $\Ex r r' = 0$, so $r$ and $r'$ are uncorrelated.
With these definitions, we define the out--of--sample error as
\beqn{equ:p40}
\eloo = \E{\eta' - y(\eta)}^2.
\eeq
By elementary manipulation
\beq{equ:p45}
\eloo = \E{\zeta - y(\eta)}^2 + s,
\eeq
so that the out--of--sample error differs from the mean square error merely by a constant $s$.
On the other hand, we have 
\beq{equ:p50}
\eloo = \E{\eta - y(\eta)}^2 + 2\Cov{\eta, y(\eta)}.
\eeq
To see this, first note that for {\em any} random variables $a, b$, 
\begin{multline*}
\E{a - b}^2 = \E{a - \bar{a}}^2 + \E{b - \bar{b}}^2 + (\bar{a} - \bar{b})^2 \\
	- 2 \Cov{a,b}.
\end{multline*}
To prove Equation~\eqref{equ:p50}, apply the previous relation to both $\E{\eta - y(\eta)}^2$ and $\E{\eta' - y(\eta)}^2$, noting that $y(\eta)$ is correlated with $\eta$ but not with $\eta'$.
The quantity $S = \Cov{\eta, y(\eta)}/s$ will be referred to as the {\em sensitivity} from now on.
The sensitivity should be understood as the correlation between $r$ filtered through the estimator $y(\eta)$, and  $r$ itself.
This correlation being large indicates that changes of the input will cause large changes of the output, while a low correlation indicates the opposite; hence the term ``sensitivity''.
We will now use the out--of--sample error to determine $\alpha$ for the specific estimator $y(\eta)$ as defined in Equation~\eqref{equ:p25}.
In the present situation, the second term in Equation~\eqref{equ:p50} becomes
\beqn{equ:p60}
2\Cov{\eta, y(\eta)} = 2\alpha s.
\eeq
So far, these have been exact calculations, but now we apply an approximation to the first term in Equation~\eqref{equ:p50}; we simply replace it by $(\eta - y(\eta))^2$.
Equation~\eqref{equ:p50} thus becomes
\beqn{equ:p70}
\E{\eta' - y(\eta)}^2
\cong (1 - \alpha)^2 (\eta - \eta_0)^2 + 2 \alpha s.
\eeq
This expression is minimal for
\beq{equ:p80}
\alpha = \frac{(\eta - \eta_0)^2 - s}{(\eta - \eta_0)^2}.
\eeq
Besides the noise strength $s$, this expression contains only available quantities.
Hence, given a rough estimate of $s$, this expression can be used as a guide to set $\alpha$ appropriately.
Expression~\eqref{equ:p80} should be compared with the optimal $\alpha$ in~\eqref{equ:p35}; within the same approximation that we did above, $(\eta - \eta_0)^2 \cong s + (\zeta - \eta_0)^2$, so that 
\beqn{equ:p85}
\alpha = \frac{(\eta - \eta_0)^2 - s}{(\eta - \eta_0)^2}
 \cong \frac{(\zeta - \eta_0)^2}{s + (\zeta - \eta_0)^2}
\eeq
That is, within our approximation, we obtain the optimal $\alpha$ (Equ.~\ref{equ:p35}) for the {\em true} mean square error.
If, alternatively, $\eta_0$ is interpreted as another observation with mean square error $s_0$, then
\beqn{equ:p85}
\frac{\alpha}{1 - \alpha} = \frac{(\eta - \eta_0)^2 - s}{s}
 \cong \frac{(\zeta - \eta_0)^2}{s}
 \cong \frac{s_0}{s},
\eeq
Hence the $\alpha$ we have determined by minimising the out--of--sample error is an approximation to the optimal $\alpha$ for the BLUE estimator.
The optimal $\alpha$ can be interpreted as an optimal weighting between the information provided by our first guess $\eta_0$ and the noise polluted observation $\eta$.
The main aim of this work is to extend these concepts to data assimilation problems.
Equation~\eqref{equ:p50} in particular will play a central role in this contribution.
Since data assimilation is dynamical in character though, the error functional should take this into account.
Hence there should be terms in Equation~\eqref{equ:p20} that reflect our first guess that $\eta$ is the result of some dynamical process; in particular, $\eta$ and $y$ are no longer just scalars, but pieces of trajectories.  
Our error functional will therefore be more complicated than in Equation~\eqref{equ:p20}.
Furthermore, our estimators will be nonlinear, which necessitates further approximations.
\subsection{Data assimilation}
\label{subsec:problem-formulation}
We suppose that a time series $\{\eta_t, t \in [t_s, t_f]\}$, referred to as the {\em observations}, has been recorded, where $\eta_t \in \R^d$ for all $t \in [t_s, t_f]$.
We will often write $\eta$ (without time index) as an abbreviation for the entire time series $\{\eta_t, t \in [t_s, t_f]\}$ of observations, and similarly for other time series.
The time interval will always be $[t_s, t_f]$, unless explicitely stated otherwise.
Data assimilation is a procedure by which trajectories $\{x_t \in \R^D, t \in [t_s, t_f]\}$ are computed which are orbits of some dynamical model, at least up to some degree of accuracy.
The exact form of the model or the assimilation procedure is not important at the moment.
Furthermore, the orbits should reproduce the observations in the following sense: There is a function $C: \R^D \to \R^d$ (which can be considered part of the model) so that the output $y_t = C(x_t)$ is close to $\eta_t$ up to some degree of accuracy.
In order to keep things simple, we will continue to work with the mean square error as a measure of closeness.
That is, we measure the deviation of the output $y_t$ from the observations  $\eta_t$ by means of the {\em tracking error}
\beq{equ:85}
A_{\mathrm{T}} := \int_{t_s}^{t_f} (\eta_t - y_t)^T W (\eta_t - y_t) \dd t,
\eeq
where $W$ is some positive definite matrix.
Unless the model is perfect or at least allows to shadow the true dynamics for long times, data assimilation will have to balance between finding a trajectory which is close to the observations, and finding a trajectory which is close to being an orbit of the model.
In this section, a criterion is proposed which allows to settle the trade--off.
The first assumption we need in order to render the criterion applicable is that the data assimilation under concern is able to explore the trade--off.
More specifically, we assume that there is a parameter $\alpha \in [0, 1]$ so that when $\alpha = 0$, the data assimilation produces trajectories which are orbits of the model while paying minimal attention to the observations.
When $\alpha = 1$, then the data assimilation is assumed to follow the observations as close as possible, while paying minimal attention to the model dynamics.
This parameter will be referred to as the {\em sensitivity parameter} from now on, for reasons that will become clear later.
Arguably, any data assimilation should possess such a knob, even though it is often hidden behind the fascia of the algorithm.
The specific approaches to data assimilation considered later should illustrate this point.
In Subsection~\ref{subsec:prelude}, we encountered a trivial version of this trade off in which the first guess $\eta_0$ played the role of the ``model'' for the observations.
There, the knob for the trade--off was given by $\alpha$.
In order to distinguish points on the trade--off, we consider some form of out--of--sample error.
For every $t \in [t_s, t_f]$, data assimilation provides an operator $\y_t$ which maps the entire time series of observations $\eta$ onto $y_t = \y_t(\eta)$.
In fact, the operators $\y_t$ depend also on the sensitivity parameter $\alpha$, which will become important later.
Until then, the dependence on $\alpha$ will not be made explicit, in order to avoid notational clutter.
Note that the role of $Y_t$ played here is analogous to that of $y(\eta)$ in Subsection~\ref{subsec:prelude}. 
Roughly speaking, we want the operators $\y_t$ to separate $\eta$ into a ``desired'' and an ``undesired'' part.
To this end, we assume that $\eta_t$ can be written as $\eta_t = \zeta_t + r_t$ for all $t \in [t_s, t_f]$, where $\zeta_t$ is the desired and an $r_t$ the undesired part.
The result of the data assimilation can now be written as $y_t = \y_t(\zeta + r)$ for all $t \in [t_s, t_f]$.
To define the out--of--sample error, we assume that $r$, henceforth called the {\em observational noise}, is some stochastic process with zero mean, and that $r$ and $\zeta$ are uncorrelated.
(This does not necessarily mean that $r$ is a white noise process, nor that it is particularly irregular.)
Now let $\eta'_t = \zeta_t + r'_t$, where $r'$ has exactly the same stochastic characteristics as $r$ but is independent from the latter.
We define the out--of--sample error as
\beq{equ:80}
\eloo := \int_{t_s}^{t_f} \E{(\eta'_t - y_t)^T W (\eta'_t - y_t)} \dd t,
\eeq
where $y_t = \y_t(\zeta + r)$ (note the absence of the dash on $r$).
The expectation $\Ex$ in Equation~\eqref{equ:80} affects both $r$ and $r'$.
The out--of--sample error can be considered as measuring some kind of universality property of our data assimilation algorithm.
Since $r'$ is uncorrelated to {\em both} $\zeta$ and $y$, we have (analogously to Equ.~\ref{equ:p45})
\beq{equ:84}
\begin{split}
\eloo
 & = \int_{t_s}^{t_f} \E{(\zeta_t - y_t)^T W (\zeta_t - y_t)} \dd t \\
 & \quad + \int_{t_s}^{t_f} \E{r_t^T W r_t} \dd t.
\end{split}
\eeq
The cross terms cancel.
The second term can be written as
\[
\int_{t_s}^{t_f} \E{r_t^T W r_t} \dd t
 = \int_{t_s}^{t_f} \tr (W \rho_0) \dd t
\]
where $\rho_0$ is the variance of the noise (which we assume to be independent of $t$; see also Equ.~\ref{equ:153}).
Hence the second term in Equation~\eqref{equ:84} depends on the noise alone and is not affected by the data assimilation.
The first term of
Equation~\eqref{equ:84},
\beqn{equ:8444}
A_\mathrm{A} := \int_{t_s}^{t_f} \E{(\zeta_t - y_t)^T W (\zeta_t -
y_t)} \dd t
\eeq 
will be referred to as the {\em assimilation error} from now
on.
The reader should note the analogy between the assimilation error $A_{\mathrm{A}}$ and the mean--square--error of Subsection~\ref{subsec:prelude}, Equation~\eqref{equ:p10}.
\subsection{Calculating the out--of--sample error}
\label{subsec:calculating-eoos}
Typically, data assimilation algorithms work by minimising some error functional which includes some form of tracking error $A_{\mathrm{T}}$ (Equ.~\ref{equ:85}) plus some hard or soft constraints which take into account the dynamical character of the problem.
Invoking some kind of stationarity though, we can reasonably hope that the variations of $A_{\mathrm{T}}$ are in fact small, that is,
\beq{equ:90}
\Ex  A_{\mathrm{T}} \cong A_{\mathrm{T}} = \int_{t_s}^{t_f} (\eta_t - y_t)^T W (\eta_t - y_t) \dd t.
\eeq
Analogously to Equation~\eqref{equ:p50} in Section~\ref{subsec:prelude}, we should not expect the tracking error to be even approximately equal to the out--of--sample error, since the integrand in $A_{\mathrm{T}}$ involves the difference $\eta_t - y_t = \zeta_t + r_t - \y_t(\zeta + r)$, while the out--of--sample error uses $\eta'_t - y_t = \zeta_t + r'_t - \y_t(\zeta + r)$;
note the distribution of dashes.
More specifically, we have
\begin{multline*}
\E{(\eta'_t - y_t)^T W (\eta'_t - y_t)} = \E{(\eta_t - y_t)^T W (\eta_t - y_t)}\\
 + 2 \tr \left( W \Cov{y_t, \eta_t} \right),
\end{multline*}
with $\bar{y}_t = \E{y_t}$.
The proof is the same as for~\eqref{equ:p50}: expand the quadratic terms on both sides and note that $y_t$ is correlated with $\eta$ but not with $\eta'$.
By integrating over time, we get
\beqn{equ:149}
\eloo = \Ex  A_{\mathrm{T}}
    + 2 \int_{t_s}^{t_f} \tr \left( W \Cov{y_t, \eta_t} \right) \dd t.
\eeq
We will write this as
\beq{equ:150}
\eloo = \Ex  A_{\mathrm{T}}
    + 2 \tr(W S \rho_0),
\eeq
with
\beq{equ:153}
\rho_0 := \Ex r_t r_t^T 
\eeq
being the variance of the noise, and
\beq{equ:155}
S := \int_{t_s}^{t_f} \Cov{y_t, \eta_t} \rho_0^{-1}\dd t,
\eeq
which we again refer to as the sensitivity.
Note that the sensitivity is essentially the integral of a correlation and thus has dimension time.
What renders Equation~\eqref{equ:150} interesting is that it, at least in principle, opens up the possibility of approximating $\eloo$ operationally.
To this end, the approximation~\eqref{equ:90} would be used for the tracking error, while for the second term, the statistics of $r$ would be required as well as the sensitivity.
In practice, an exact calculation of $S$ might be a serious difficulty given the complexity of data assimilation algorithms.
In order to get any further, we assume that the effect of $r$ on the reconstructed observations $y$ can be described through a linear analysis.
More specifically, 
\beq{equ:160}
y_t = \y_t(\zeta + r) \cong \y_t( \zeta)  + \dd \y_t (\eta) * r
\eeq
where $\dd \y_t (\eta)$ is a linear operator, describing the first order response of the operator $\y_t$ at $\eta$.
Taking the linearisation at $\eta$ rather than at $\zeta$ is maybe less customary, but the error commited by either choice is of the same order.
By $\dd \y_t (\eta) * r$, we denote the application of that linear operator to $r$.
From this assumption, we get that to first order
\beq{equ:170}
\bar{y}_t \cong \y_t(\zeta).
\eeq
Since
\[
\Cov{y_t, \eta_t} 
= \E{(y_t - \bar{y}_t)^T (\eta_t - \bar{\eta}_t)}
= \E{(y_t - \bar{y}_t)^T r_t}
\]
we obtain for the sensitivity by substituting with~(\ref{equ:160},~\ref{equ:170}) in~\eqref{equ:155}
\beq{equ:180}
S \cong \int_{t_s}^{t_f} \E{(\dd \y_t (\eta) * r) r_t^T} \rho_0^{-1} \dd t.
\eeq
This linear approximation will be at the basis of subsequent calculations of the sensitivity.
As was mentioned in the introduction, a related sensitivity concept has been studied in the literature already, for example in \citet{cardinali04,liu_estimating_2008,liu_analysis_2009}.
As far as these studies pertain to 3D--Var, the current output $y_t$ at time $t$ is assumed to be linearly regressed from the current observation $\eta_t$ and the background $x_t$ which contains all previous information.
Only the influence of the current observation  $\eta_t$ onto $y_t$ is taken into account.
The effect of previous observations on $y_t$ (via the background) is not studied.
In contrast, the sensitivity as defined in the present paper considers the influence of the entire history of observations (past and future) onto the output $y_t$.
Our original motivation for studying the out--of--sample error and the sensitivity was to get a handle on the trade--off of data assimilation.
The precise relation between the sensitivity and what was called the sensitivity parameter before obviously depends on the details of the data assimilation scheme.
It is to be expected though that, in any event, the sensitivity is intimately related to the trade--off: to create solutions that track the observations closely, a large sensitivity is needed, which however introduces dynamical errors.
Computing trajectories which adhere to the model dynamics on the other hand is only possible with a low sensitivity.
In order to compute $S$ in practice, firstly the linearisations of the operators $\y_t$ about $\eta$ have to be calculated.
These are often already computed as part of the data assimilation procedure itself, in which case this typically rather burdensome task need not be performed twice.
Again, the details depend on the specific data assimilation technique.
Two general approaches to data assimilation will be considered in the next two sections, and ways to compute the sensitivity will be discussed.
The correlation structure of $r$ has to be known at least to some degree.
In fact, there is nothing really to be known here, but rather a decision has to be made which parts of the observations are to be considered of interest and which parts are to be considered noise.
In the next subsection, an approach to this problem is discussed which hopefully covers a large range of applications.
%
%
%
%
%%%%%%%%%%%%%%%%%%%%%%%%%%%%%%%%%%%%%%%%%%%%%
%
\subsection{Assumptions on the noise $r$}
\label{subsec:noise-assumptions}
The following assumptions on $r$, which typically apply in practical situations, allow to simplify the calculation of the sensitivity in the cases considered in this paper.
\begin{enumerate}
\item \label{noise-assumption-a} $r$ is sampled from a signal $\nu$ with sampling interval $\Delta t$ and subsequently interpolated.
\item \label{noise-assumption-b} $\nu$ is stationary in the wide sense.
\item \label{noise-assumption-c} $\frac{1}{\Delta t}$ is small with respect to the bandwidth of $\nu$.
\end{enumerate}
It follows from assumption~\ref{noise-assumption-b} that the covariance function $\Ex{\nu_t\nu^T_s}$ depends only on $t-s$.
Furthermore, $\nu$ has a well defined power spectrum.
Now let
\[
\rho_{ts} := \E{r_t r_s^T}.
\]
It follows from assumption~\ref{noise-assumption-a} that also $\rho_{t, s}$ depends on $t - s$ only, and that $r$ has a well defined power spectrum.
This power spectrum is confined to a band of width $\frac{1}{\Delta t}$.
Furthermore, during sampling, all power of $\nu$ beyond the critical frequency $\frac{1}{2\Delta t}$ will be aliased into that band.
Hence and from~\ref{noise-assumption-c}, it follows that $r$ has still power near the critical frequency $\frac{1}{2\Delta t}$.
In summary, we can conclude that the power spectral density of $r$ can, in good approximation, be written as
\beq{equ:400}
\rho^*(f) \cong
\begin{cases} \rho_0 \Delta t & \text{if } |f| \leq \frac{1}{2\Delta t},\\
0 & \text{otherwise.}
\end{cases}
\eeq
In effect, we assume $r_t$ to be band limited white noise.
For a discussion how band limited white noise arises through sampling white noise, see~\citet{astrom_stochastic_control_2006}.
If there is good reason to believe that assumption~\ref{noise-assumption-c} does not hold, and $\frac{1}{\Delta t}$ is on the same order of magnitude as the bandwidth of $\nu$ or larger, then Equation~\eqref{equ:400} might still be a reasonable approximation, provided $\Delta t$ is replaced by the bandwidth of $\nu$.
In this paper, we will in fact not employ the specific form of the correlation function~\eqref{equ:400}.
The only thing we require is the validity of the following two approximations~\citep[see][Ch.~2, Sec.~5]{astrom_stochastic_control_2006}.
Suppose that $\phi$ is a function on some interval $[t_1, t_2]$ which varies slowly compared to $\Delta t$.
If $t$ is well inside the interval $[t_1, t_2]$, then
\beq{equ:slowly-var}
\int_{t_1}^{t_2} \phi_{\tau} \rho_{\tau - t} \dd \tau \cong \rho_0 \Delta t \phi_{t}
\eeq
If however $t = t_2$, then due to the symmetry of $\rho$,
\beq{equ:slowly-var-2}
\int_{t_1}^{t_2} \phi_{\tau} \rho_{\tau - t} \dd \tau \cong \frac{1}{2} \rho_0 \Delta t \phi_{t}.
\eeq
\section{Example 1: Variational data assimilation}
\label{sec:variational-data-assimilation}
In this section, a class of data assimilation approaches, often referred to as variational data assimilation, will be considered.
After recalling the basics of this approach, we will indentify the sensitivity parameter $\alpha$.
In order to set $\alpha$ with the help of the out--of--sample error, we need to calculate the sensitivity.
This will essentially occupy the present section.
We assume a model of the form
\beq{equ:10}
\dot{x}_t = f(x_t) + u_t, \qquad y_t = Cx_t
\eeq
with {\em state} $x_t \in \R^D$ and {\em output} $y_t \in \R^d$.
The {\em model dynamics} $f$ is a vector field on some open subset of $\R^D$, $C$ is a $d \times D$--matrix.
The model dynamics should be thought of as capturing our a priori knowledge of the physical mechanisms underlying the observations $\eta$.
For this reason, the $u_t \in \R^D$ will be referred to as the {\em dynamical perturbations}.
{\em Variational Data Assimilation} attempts to find state and dynamical perturbation trajectories $\{(x_t, u_t), t \in [t_s, t_f]\}$ satisfying the relations~\eqref{equ:10} so that the {\em action integral}
\begin{multline}
\label{equ:20}
A_{\alpha}\{x, u\}
    := \frac{\alpha}{2} \int_{t_s}^{t_f}
        (\eta_t - y_t)^T R (\eta_t - y_t)\dd t \\
    + \frac{1 - \alpha}{2} \int_{t_s}^{t_f} u_t^T Q u_t \dd t
\end{multline}
is minimal.
Here, $R$ and $Q$ are positive definite matrices which might be necessary to appropriately scale different degrees of freedom.
This approach has been studied in various publications, see for example~\citet{ledimet86,courtier87,derber89,apte08}.
We shall now digress for two paragraphs and discuss the various arguments which have been put forward as to why an action integral of the form~\eqref{equ:30} should be used.
In the case of linear dynamics, the variational approach (to be described below) in conjunction with the action integral~\eqref{equ:20} leads, somewhat coincidentially, to the Kalman filter and smoother.
This fact, which is a consequence of Kalman's duality theorem, can be seen as an extension of the BLUE concept mentioned in Subsecion~\ref{subsec:prelude}~\citep[see][]{JAZ,sage68}.
These facts do no longer hold if the dynamics are nonlinear.
Yet an alternative interpretation of the action integral is as the logarithmic aposteriori density of a trajectory~\citep{JAZ}, given the observations.
In connection with this interpretation though, it is necessary to keep the following reservations in mind.
Firstly, the perturbations to the dynamics and the observations need to be gaussian and uncorrelated in order for this interpretation to apply.
Secondly, $R$ and $Q$ have to be the inverse covariance matrices of the observational and the dynamical perturbations, respectively (and $\alpha = 1/2$); otherwise, the functional we minimise is not the logarithmic aposteriori.
Thirdly, in a nonlinear context, the maximum--aposteriori estimator does not necessarily yield a minimum mean square error.
Finally, the very concept of a ``density'' requires substantial modifications in continuous time; \citet{zeitouni_maximum_1987} discuss an aposteriori which, in general, looks different from Equation~\eqref{equ:20}.
(The numerical examples we will discuss later happen to be an exception; see the discussion in Sec.~\ref{sec:numerical-experiments}.)
Without invoking statistical concepts, we contend that an error functional of the form~\eqref{equ:20} might still be justified for the purpose of regularisation.
For a general discussion and motivation of this view (adopted here), we refer the reader to~\citet{broecker_variational_2010}.
The parameter $\alpha$, which is analogous to the parameter $\alpha$ in Subsection~\ref{subsec:prelude}, is interpreted a regularisation parameter. 
(It is fair to say that under this paradigm, there is no a~priori reason to use a quadratic measure of error, which is used here mainly for mathematical convenience.)
Whichever view on the interpretation of the functional $A_{\alpha}$ is adopted, there remains the problem of setting $\alpha$.
The parameter $\alpha$ controls the weighting between the two contributions to the action integral, namely between the tracking error
\beqn{equ:60}
A_{\mathrm{T}} = \frac{1}{2} \int_{t_s}^{t_f} (\eta_t - y_t)^T R (\eta_t - y_t) \dd t
\eeq
and the {\em model error}
\beqn{equ:70}
A_{\mathrm{M}} = \frac{1}{2} \int_{t_s}^{t_f} u_t^T Q
u_t \dd t. 
\eeq
Note that the tracking error was already defined as a diagnostic in Equation~\eqref{equ:85}, albeit with a weighting matrix called~$W$.
That tracking error might be referred to as the {\em diagnostic} tracking error.
There is no need though for using the same weighting matrix in the diagnostic tracking error and in the functional $A$, whence we will keep them distinct.
For the variational approach, $\alpha$ is a sensitivity parameter in accordance with the definition given in Subsection~\ref{subsec:problem-formulation}.
If $\alpha$ is close to one, there is almost no penalty on the dynamical perturbations.
The problem of minimising the action integral then becomes a very easy one, as arbitrary dynamical perturbations can be used to make the tracking error small.
The solution then follows the observations very closely, but in general it will not be a good solution of the model dynamics, as $u_t$ will be large.
If $\alpha$ is close to zero, there is almost no penalty on the observations.
\citep[The problem of minimising the action integral in this situation is generally not very easy, since the problem tends to have a very poor condition; nonetheless, acceptable solutions for small $\alpha$ can be found through {\em continuation}, see][]{broecker_variational_2010}.
The solution will be a good solution of the model dynamics in the sense that $u_t$ is small in this situation, but in general there will be deviations from the observations.
In summary, varying $\alpha$ allows for exploring the trade--off between finding a trajectory which is close to the observations, and finding a trajectory which is close to being an orbit of the model.
In principle, the matrices $R, S$ can be considered an entire set of sensitivity parameters which could be determined by the approach proposed in this paper.
In the examples studied in this paper though, we will work with fixed matrices $R, S$ and study the dependence on the scalar sensitivity parameter $\alpha$ only.
To compute the out--of--sample error by means of the sensitivity, some remarks are necessary as to how to solve the variational data assimilation problem.
The following results are classical, see for example~\citet{sage68}.
A variational problem under constraints can be transformed into a variational problem without constraints by introducing dual variables or Lagrange multipliers.
More specifically, the problem above is equivalent to finding a stationary point (without constraints) of the {\em augmented action}
\begin{multline}
\label{equ:30}
\bar{A}_{\alpha}\{x_t, u_t, \lambda_t, \eta_t\}
    := \int_{t_s}^{t_f} H_{\alpha}(x_t, u_t, \lambda_t, \eta_t) \dd t \\
    - \int_{t_s}^{t_f} \lambda_t^T \dot{x}_t \dd t
\end{multline}
over $\{(x_t, u_t, \lambda_t), t \in [t_s, t_f]\}$ for a fixed time series $\eta$, where the {\em Hamiltonian} is given by
\begin{multline*}
% \label{equ:40}
H_{\alpha}(x, u, \lambda, \eta)
    = \frac{\alpha}{2} (\eta - Cx)^T R (\eta - Cx) \\
    + \frac{1 - \alpha}{2} u^T Q u + \lambda^T \left( f(x) + u \right).
\end{multline*}
Since no derivatives of $u_t$ appear, the minimisation over $u_t$ can be immediately effected, leading to the criterion
\[
\partial_{u} H_{\alpha} = 0 \qquad \mbox{for all $t$},
\]
which gives 
\beqn{equ:35}
u_t = - \frac{1}{1 - \alpha} Q^{-1} \lambda_t.
\eeq
Substituting with this expression for $u$, we obtain
\begin{multline}
\label{equ:40}
H_{\alpha}(x, \lambda, \eta)
    = \frac{\alpha}{2} (\eta - Cx)^T R (\eta - Cx) \\
    - \frac{1}{2 (1 - \alpha)} \lambda^T Q^{-1} \lambda + \lambda^T f(x).
\end{multline}
as the definitive expression for the Hamiltonian.
The following equations are necessary and sufficient conditions for $\{(x, \lambda)\}$ to be a stationary trajectory of the augmented action~\eqref{equ:30}:
\begin{align}
\partial_{\lambda} H_{\alpha} - \dot{x} & = 0,
    \label{equ:50.1}\\
\partial_x H_{\alpha} + \dot{\lambda} & = 0,
    \label{equ:50.2}\\
\lambda_{t_s} = 0, & \quad \lambda_{t_f} = 0.
    \label{equ:50.4}
\end{align}
The Equations~(\ref{equ:50.1},\ref{equ:50.2}) describe the dynamical evolution of the states $x$ and the co--states $\lambda$.
The conditions in Equation~\eqref{equ:50.4} represent boundary conditions.
Depending on the specific circumstances, other boundary conditions might be appropriate.
For example, imposing (hard or soft) initial or terminal conditions on $x_t$ in the problem formulation (``background error'') would lead to modifications in the boundary conditions~\eqref{equ:50.4}.
Such modifications though do not change the general character of the problem.
To keep this discussion as short as possible, we will work with the simple boundary conditions~\eqref{equ:50.4}.
For the specific Hamiltonian~\eqref{equ:40} the necessary conditions~(\ref{equ:50.1}--\ref{equ:50.4}) read as
\begin{align}
\dot{x}_t & = f(x_t) - \frac{1}{1 - \alpha} Q^{-1} \lambda_t,
    \label{equ:55.1}\\
\dot{\lambda}_t & =  - Df(x_t)^T \lambda_t + \alpha C^T R (\eta_t - C x_t),
    \label{equ:55.2}\\
\lambda_{t_s} & = 0, \quad \lambda_{t_f} = 0.
    \label{equ:55.4}
\end{align}
The same formalism can be applied to much more general setups, such as more general integrants in the action as well as more general dynamical systems.
With the appropriate Hamiltonian $H$, the necessary conditions for an optimum always look like Equations~(\ref{equ:50.1},\ref{equ:50.2}).
Instead of an initial value problem where the state $(x_t, \lambda_t)$ is specified at $t = t_s$, we face a {\em two--point boundary value problem} or~BVP for short\footnote{All boundary value problems in this paper are two--point, so using the general abbreviation BVP should not give rise to confusion}, in which  the state is specified partly at the initial time and partly at the terminal time.
BVP's require other solution algorithms than initial value problems, and a large variety of numerical techniques have been developed.
Our aim though is to solve not only the Hamiltonian BVP but also to calculate the sensitivity.
Therefore, in order to save computational resources, quantities that have already been calculated during the BVP solving should be reused as much as possible.
Considering every conceivable BVP solver and demonstrating how it could be extended to yield also the sensitivity is obviously beyond the scope of this paper.
Discussing two general approaches to BVP solving will have to be sufficient.
We hope that these two examples are general enough to be transferable to more specialised BVP solving approaches.
A very general method to solving BVP's is the collocation method, considered in Subsection~\ref{subsec:sensitivity-vda-colloc-jac}.
In this approach, the BVP is approximated on a discrete time mesh by a set of nonlinear equations, the so--called {\em collocation equations}.
The sensitivity is related to the Jacobian of the collocation equations at the optimum.
If the Jacobian calculated during the solution of the collocation equations can be recycled, then the sensitivity is obtained with close to no additional computational effort.
Subsection~\ref{subsec:sensitivity-vda-colloc-jac} also contains further references on this topic.
In the collocation approach, the entire set of collocation equations is solved simultaneously.
This renders the collocation approach numerically expensive for large (e.g.\ spatially extended) systems, even if the fact is exploited that the linear approximation is typically a sparse matrix.
In these situations, sequential approaches are presumably more economical.
Subsection~\ref{subsec:sensitivity-vda-bvp} considers such an approach.
It is based on the fact that the linear response of the Hamiltonian BVP (with respect to perturbations of the inputs) can be described by a linear Hamiltonian~BVP.
The solution of the latter BVP (and subsequently, the sensitivity) is obtained by consecutively solving matrix valued differential equations of Ricatti type.
%
%
%
%
%%%%%%%%%%%%%%%%%%%%%%%%%%%%%%%%%%%%%%%%%%%%%
%
%
\subsection{The sensitivity through the Jacobian of the collocation equations}
\label{subsec:sensitivity-vda-colloc-jac}
We start with explaining the general idea of the collocation method.
For more details, see for example \citet{kierzenka01}, which contains a discussion of the \texttt{bvp4c}--algorithm implemented in Matlab that uses the collocation method.
For the moment, we are not using the fact that the vector field is Hamiltonian, so we will be assuming (until further notice) that the BVP is given as
\beq{equ:265}
\dot{z}_t = F(t, z_t), \qquad t \in [t_s, t_f], \qquad b(z_{t_s}, z_{t_f}) = 0,
\eeq
with $z_t = (x_t,\lambda_t) \in \R^{2D}$, $F$ a vector field on
$\R^{2D}$ and $b(., ..)$ a function representing the boundary
conditions.
The general strategy of most BVP solvers is to approximate $z_t$ at a series $t_s = t_0 < \ldots < t_N = t_f$ of temporal mesh points; we will write $\z := (z_{t_0}, \ldots, z_{t_N})$ for the values of the solution at these points.
Usually, $\z$ is obtained by solving a set of equations
\beqn{equ:270}
\Phi_i(\z) = 0, \qquad i = 0 \ldots N,
\eeq
called {\em collocation equations}.
The collocation equations are effectively discrete time approximations of the original differential equation as well as the boundary conditions in~\eqref{equ:265}.
More specifically, the $\Phi_i$ are functions of $F_i = F(t_i,
z_{t_i})$ and the boundary conditions $b(z_{t_0}, z_{t_N})$.
For the BVP solver \texttt{bvp4c}~by \citet{kierzenka01} implemented in Matlab, the explicit form of $\Phi$ is given in the Appendix.
Typically, the collocation equations are solved using a quasi--Newton type algorithm, which involves a numerical approximation to the Jacobian of the collocation equations.
%
%(Alternatively, the BVP solver might use analytical Jacobians of the vector field and boundary conditions to assemble the Jacobian of the collocation equations, if these data are provided by the user.)
%

%
Once available, the Jacobian can be re--used to compute the sensitivity, as will be explained now.
As was explained in Subsection~\ref{subsec:sensitivity-vda-bvp}, a  perturbation $r$ added to the observations will result in a perturbation of the Hamiltonian vector field (which we have written as $F$ in this subsection), which in turn will entail perturbations of the solution (which we have written as $\z$ in this subsection).
In terms of the collocation equations, it means that we have a slightly perturbed solution $\z + \delta \z$ of some slightly perturbed collocation equation
\beqn{equ:300}
(\Phi + \delta\Phi) (\z + \delta \z) = 0.
\eeq
Expanding the left hand side and keeping only terms up to linear order in the small quantities $\delta\Phi$ and $\delta\z$, we obtain
\beqn{equ:310}
0 = \Phi(\z) + D\Phi(\z)\delta\z + \delta\Phi(\z).
\eeq
The first term vanishes since $\z$ is a solution of the unperturbed collocation equations, per assumption.
Solving for $\delta\z$ gives
\beq{equ:320}
\delta\z = - D\Phi(\z)^{-1} \delta\Phi(\z).
\eeq
Since the perturbation $\delta\Phi(\z)$ of the collocation equations is due to perturbations $\delta F$ of the vector field, we have to first order
\beq{equ:330}
\delta\Phi_i(\z) = \sum_j \frac{\partial \Phi_i}{\partial F_j}(\z)\delta F_j,
\eeq
It remains to express $\delta F_j$ in terms of $r$.
To this end, we have to resort to the particular form of $F$ given in Equations~\eqref{equ:55.1} and~\eqref{equ:55.2}.
It follows that
\beq{equ:370}
\delta F_j = \left( \begin{array}{c} 0 \\ C^T R r_{t_j} \end{array}\right).
\eeq
Therefore, by combining Equations~\eqref{equ:320},~\eqref{equ:330}, and~\eqref{equ:370}, along with the fact that $\dd \y_{t_i} = \left(C, 0\right) \delta z_i$, we get that
\beq{equ:380}
\dd \y_{t_i} = -\sum_{j,k} \left(C, 0\right) \cdot D\Phi_{ij}^{-1} \frac{\partial \Phi_j}{\partial F_k}(\z) \cdot \left( \begin{array}{c} 0 \\ C^T R r_{t_k} \end{array}\right).
\eeq
For the correlation $\E{ (\dd \y_{t_i} * r) r_{t_i}^T}$ we get from Equation~\eqref{equ:380}
\beqn{equ:390}
\begin{split}
\lefteqn{\E{ (\dd \y_{t_i} * r) r_{t_i}^T}} & \\
 & = -\E{ \sum_{j,k} \left(C, 0\right) \cdot D\Phi_{ij}^{-1}
	\frac{\partial\Phi_j}{\partial F_k}(\z)
	\cdot 
	\left( \begin{array}{c} 0 \\
		\alpha C^T R r_{t_i}\end{array}
	\right)
	\cdot r^T_{t_l}} \\
 & = -\sum_{j,k} \left(C, 0\right) \cdot D\Phi_{ij}^{-1}
	\frac{\partial\Phi_j}{\partial F_k}(\z)
	\cdot 
	\left( \begin{array}{c} 0 \\
		\alpha C^T R \rho_{t_k - t_i} \end{array}
	\right)		
\end{split}
\eeq
Using the trapezoidal rule over the time mesh of the BVP solver, we obtain the following approximation to the sensitivity:
\beq{equ:395}
\begin{split}
S & = -\sum_{i,j,k} h_{i-1} \left(C, 0\right) \cdot D\Phi_{ij}^{-1} 
	\frac{\partial \Phi_j}{\partial F_k}(\z) 	\\
	& \quad \cdot \times \cdot 
	\left( \begin{array}{c} 0 \\
		\alpha C^T R \rho_{t_k - t_i} \end{array}
	\right) \rho^{-1}_0
\end{split}
\eeq
with $h_{i} = t_{i+1} - t_{i}$.
Computing the sensitivity through this relation becomes attractive if the inverse $D\Phi_{ij}^{-1}$ of the collocation Jacobian is already available from the BVP solver, as this is the most costly step in evaluating Equation~\eqref{equ:395}.
Secondly, the partial derivatives $\partial \Phi_i / \partial F_j$ can often be extracted from the BVP solver as well.
The explicit form of the partial derivatives $\partial \Phi_i / \partial F_j$ for the BVP solver \texttt{bvp4c} is given in the Appendix.
Equation~\eqref{equ:395} is valid without imposing any specific conditions on the covariance $\rho_{t,s}$ of the signal $r$.
Under the additional assumptions as outlined in Subsection~\ref{subsec:noise-assumptions}, many of the off--diagonal terms in the sum over $i, k$ of Equation~\eqref{equ:395} will be zero, which might considerably reduce the necessary amount of calculations.
%
%
%%%%%%%%%%%%%%%%%%%%%%%%%%%%%%%%%%%%%%%%%%%%%
%
%
\subsection{The sensitivity through solution of a linear BVP}
\label{subsec:sensitivity-vda-bvp}
The general strategy of solving nonlinear problems by solving a series of linear problems which approximate the original problem is, in principle, also applicable to BVP's.
If such a strategy is employed, then Equations of the form~(\ref{equ:190.1},\ref{equ:190.2}) below get solved.
This means that all essential calculations of the sensitivity calculations below have already been carried out and can be recycled.
Suppose that $\{x, \lambda\}$ is a solution of the BVP~(\ref{equ:50.1}-\ref{equ:50.4}) with $\eta = \zeta$.
A perturbation $r_t$ added to the observations will result in a perturbation of the Hamiltonian vector field, which in turn will entail perturbations $(\xi_t, l_t)$ of the original solution $(x_t, \lambda_t)$.
To first order, $\xi_t$  and $l_t$ are given by linearisation of the Equations~(\ref{equ:50.1}-\ref{equ:50.4}) about $\{x, \lambda_t\}$, which read as
\begin{align}
\label{equ:190.1}
\dot{\xi}_t & = F_t \xi_t + M_t l_t, \\
\label{equ:190.2}
\dot{l}_t & = -N_t \xi_t - F^T_t l_t - D_t r_t,
\end{align}
with the identification
\begin{align*}
F_t & = \partial_{x\lambda} H(x_t, \lambda_t, \zeta_t) \\% \label{equ:200.1}\\
M_t & = \partial_{\lambda^2} H(x_t, \lambda_t, \zeta_t) \\%  \label{equ:200.2}\\
N_t & = \partial_{x^2} H(x_t, \lambda_t, \zeta_t)  \\% \label{equ:200.3}\\
D_t & = \partial_{\eta x} H(x_t, \lambda_t, \zeta_t) % \label{equ:200.4}.
\end{align*}
We assume that $\partial_{\eta \lambda} H = 0$, as is the case for the specific Hamiltonian~\eqref{equ:40}.
The boundary conditions for Equations~(\ref{equ:190.1}-\ref{equ:190.2}) are $l_{t_s} = l_{t_f} = 0$.
Having solved the Equations~(\ref{equ:190.1}-\ref{equ:190.2}), the linear response is given by $\dd \y_t (\zeta) * r = C \xi_t$, whence the sensitivity can be written as
\beq{equ:205}
S = \int_{t_s}^{t_f} \E{C \xi_t r^T_t } \dd t \rho^{-1}_0
 =  C \int_{t_s}^{t_f} \E{\xi_t r^T_t } \dd t \rho^{-1}_0
\eeq
We therefore compute the correlation $\Gamma_t := \E{\xi_t r^T_t}$.
It is possible to decouple Equations~(\ref{equ:190.1},\ref{equ:190.2}) by using {\em invariant imbedding}~\citep[see e.g.][]{sage68}.
In the present case, this means to try the approach $l_t = P_t \xi_t + \mu_t$.
Substituting with this approach for $l_t$ in~(\ref{equ:190.1}-\ref{equ:190.2}) and equating like powers in $\xi_t$, we obtain
\begin{align}
\label{equ:210.1}
\dot{\xi}_t & = B_t \xi_t + M_t \mu_t \\
\label{equ:210.2}
\dot{\mu}_t & = - B^T_t \mu_t - D_t r_t \\
\label{equ:210.3}
B_t & := F_t + M_t P_t \\
\label{equ:210.4}
-\dot{P}_t & = P_t F_t + F^T_t P_t + P_t M_t P_t + N_t,
\end{align}
with conditions $\mu_{t_f} = 0$, $P_{t_f} = 0$, $\xi_{t_s} = - P^{-1}_{t_s} \mu_{t_s}$.
To solve this system, Equation~\eqref{equ:210.4} is integrated first backward in time, simultaneously with Equation~\eqref{equ:210.2}.
This is possible since these Equations do not depend on $\xi$ and the end conditions are given.
Then the Equation~\eqref{equ:210.1} is solved forward in time.
The solution $l_t$ of~\eqref{equ:190.2} can be recovered through $l_t = P_t \xi_t + \mu_t$.
The solutions $\xi_t$ and $\mu_t$ can be written as
\begin{align}
\label{equ:220.1}
\xi_t & = \phi_t \left[ - \phi^{-1}_{t_s} P^{-1}_{t_s} \mu_{t_s}
    + \int^{t}_{t_s} \phi^{-1}_{s} M_s \mu_s \dd s \right] \\
\label{equ:220.2}
\mu_t & = \phi^{-T}_t \int^{t_f}_{t} \phi^{T}_{s} D_s r_s \dd s,
\end{align}
with $\phi_t$ a fundamental system of
\[
\dot{\phi}_t = B_t \phi_t.
\]
Hence, multiplying Equation~\eqref{equ:220.1} with $r^T_t$ from the right and taking $\E{\ldots}$ gives
\beq{equ:230}
\Gamma_t = \phi_t \left[ - \phi^{-1}_{t_s} P^{-1}_{t_s} \E{\mu_{t_s} r^{T}_t}
    + \int^{t}_{t_s} \phi^{-1}_{s} M_s \E{\mu_s r^{T}_t} \dd s \right].
\eeq
For $s \leq t$, multiplying Equation~\eqref{equ:220.2} with $r^T_t$ from the right and taking $\E{\ldots}$ we get
\beqn{equ:240}
\begin{split}
\E{\mu_s r^{T}_t} & = \phi^{-T}_s \int^{t_f}_{s} \phi^{T}_{\tau} D_{\tau} \E{r_{\tau} r^{T}_t} \dd \tau \\
 & = \phi^{-T}_s \int^{t_f}_{s} \phi^{T}_{\tau} D_{\tau} \rho_{\tau - t} \dd \tau.
\end{split}
\eeq
As to the correlation structure of $r_t$, we again impose the conditions of Subsection~\ref{subsec:noise-assumptions}.
In particular, we assume that $\phi^{T}_{\tau} D_{\tau}$ varies slowly compared to $\Delta t$.
Thus, Equation~\eqref{equ:slowly-var} can be applied, and we get
\begin{align}
\E{\mu_s r^{T}_t}
    & = \phi^{-T}_s \int^{t_f}_{s} \phi^{T}_{\tau} D_{\tau} \rho_{\tau - t} \dd \tau   \nonumber \\
    & \cong \Delta t \phi^{-T}_s \phi^{T}_{t} D_{t} \rho_0. \label{equ:245}
\end{align}
Replacing with Equation~\eqref{equ:245} in Equation~\eqref{equ:230} we obtain
\beqn{equ:250}
\Gamma_t =  \Delta t \, \phi_t
    \left[ \int^{t}_{t_s} \phi^{-1}_{s} M_s \phi^{-T}_{s} \dd s - \phi^{-1}_{t_s} P^{-1}_{t_s} \phi^{-T}_{t_s}\right]
     \phi^{T}_t D_t \rho_0.
\eeq
In fact, this can be written as $\Gamma_t =  \Delta t \, J_t D_t \rho_0$ with $J_t$ obeying the matrix valued differential equation
\beq{equ:260}
\dot{J}_t = B_t J_t + J_t B^{T}_t + M_t
\eeq
with initial condition $J_{t_s} = -P^{-1}_{t_s}$.
The sensitivity is obtained from Equation~\eqref{equ:205} as
\beq{equ:262}
S = \Delta_t C \int_{t_s}^{t_f} J_t D_t \dd t
\eeq
To summarise, in order to obtain $\Gamma_t$ we first need to linearise the Hamiltonian equations, then use this data to form the Ricatti equation~\eqref{equ:210.4}, which has to be solved backward in time.
Next, the solution $P_t$ of the Ricatti equation is used to form $B_t$, defined in Equation~\eqref{equ:210.3}.
With this data, Equation~\eqref{equ:260} for $J$ is integrated forward in time, and the sensitivity is eventually obtained from Equation~\eqref{equ:262}.
%
%
%%%%%%%%%%%%%%%%%%%%%%%%%%%%%%%%%%%%%%%%%%%%%
%
%
\section{Example II: Data assimilation through synchronisation}
\label{sec:theory-synchronisation}
Synchronisation between dynamical systems is a phenomenon which has been under study for some time, see for example~\citet{parlitz_synchronisation_99,boccaletti_synchronization_2002,pikovsky_synchronization_2003}.
As in Section~\ref{sec:variational-data-assimilation}, Equation~\eqref{equ:10}, let
\beq{equ:402}
\dot{x}_t = f(x_t) + u_t, \qquad y_t = C x_t
\eeq
a model with dynamical perturbations and output.
A {\em coupling} between the model~\eqref{equ:402} and reality is established by setting
\beqn{equ:404}
u_t = K \cdot (\eta_t - y_t),
\eeq
that is, the error between the model output and the observations is fed back into the model, with $K$ being some suitably chosen coupling matrix.
Synchronisation refers to a situation in which, due to coupling, the error $\eta_t - y_t$ becomes small asymptotically, irrespective of the initial conditions for the model.
We might then hope that $x_t$ is in some sense similar to the `true state of reality'.
This hope is supported by mathematical results stating that if the reality is indeed a dynamical system not too unsimilar to our model~\eqref{equ:400}, synchronisation occurs~\citep[see e.g.][]{pikovsky_synchronization_2003}.
Many examples for spontaneous synchronisation have been found in nature and engineering.
%
%(The term ``synchronisation'' has been used in a far more general sense in the literature.
%
%This does not mean that every newly discovered instance of synchronisation is actually a newly discovered phenomenon.)
%
In the context of data assimilation, an approach known as Newtonian nudging~\citep{seaman90} can be understood as trying to establish synchronisation between the model and some presupposed dynamics generating the observations.
More specifically, the assimilation is accomplished by integrating the following system
\beq{equ:410}
\dot{x}_t = f(x_t) + K (\eta_t - C x_t).
\eeq
The question arises how to choose an appropriate coupling.
For the experiments carried out in this paper, we set $K = \kappa \cdot C^T$, with {\em coupling parameter} $\kappa$.
Furthermore, $C$ was chosen so that $C C^T = I$ (the $d$--dimensional identity matrix).
The coupling parameter plays the same role as $\alpha$ in variational data assimilation, namely controlling the trade--off between tracking error and dynamical error.
Therefore, the coupling parameter is the sensitivity parameter in the present situation (we could set $\alpha = \frac{\kappa}{1 + \kappa}$ to obtain a sensitivity parameter between zero and one).
Indeed, from Equations~(\ref{equ:402},\ref{equ:410}) we obtain
\[
u_t = \kappa C^T (\eta_t - C x_t),
\]
which gives
\[
\frac{A_{\mathrm{M}}}{A_{\mathrm{T}}} = \kappa^2
\]
in the case of $R$ and $S$ being the identity matrix.
As in variational data assimilation, a possible criterion for choosing $\kappa$ could be a minimal out--of--sample error.
To this end, we need to calculate the sensitivity of the synchronisation approach.
The response of $x_t$ to small changes $r_t$ in the observations is, to first order, described by
\beq{equ:420}
\dot{\xi}_t = (D f(x_t) - KC)  \xi_t + K r_t
\eeq
Let $\phi_t$ be the fundamental system of the linear part, that is,
\beqn{equ:425}
\dot{\phi}_t = (D f(x_t) - KC)  \phi_t.
\eeq
The solution of~\eqref{equ:420} with $\xi_{t_s} = 0$ can be written as
\beqn{equ:430}
\xi_t = \phi_t \int_{t_s}^t \phi^{-1}_s K r_s \, \dd s
\eeq
from which we get
\beq{equ:440}
\Gamma_t
 = \E{\xi_t r^T_{t}}
 = \phi_t \int_{t_s}^t \phi^{-1}_s K \rho_{s - t} \, \dd s
\cong \frac{1}{2} K \rho_0 \Delta t
\eeq
Again, we have assumed the correlation structure $\rho_{\tau - t}$ for $r$ as discussed in Subsection~\ref{subsec:noise-assumptions} and also that $\phi^{-1}_s $ is slowly varying compared to $\Delta t$, whence Equation~\eqref{equ:slowly-var-2} applies.
Eventually, the sensitivity is obtained as
\beqn{equ:445}
\begin{split}
S & = C \int_{t_s}^{t_f} \Gamma_t \dd t \rho^{-1}_0
= \frac{\Delta t}{2} (t_f - t_s) C K \\
& = \frac{\Delta t}{2} (t_f - t_s) \kappa C C^T
\end{split}
\eeq
%
%
%%%%%%%%%%%%%%%%%%%%%%%%%%%%%%%%%%%%%%%%%%%%%
%
%
%
\section{Numerical experiments}
\label{sec:numerical-experiments}
Several numerical experiments were performed, with the aim of testing the methodology, in particular the equivalence of the approaches presented in Section~\ref{sec:variational-data-assimilation} and the validity of the linear approximation~\eqref{equ:180}.
Detailed experiments were carried out using the Lorenz'96 system, which will be discussed in more detail here.
The following experimental setup was used: The ``reality'' is given by
the Lorenz'96 system with two scales~\citep{lorenz96}, described by the
following equations:
\begin{align}
\dot{X}_i & = -X_{i-1} (X_{i-2} -X_{i+1})
-X_{i}+ F-\gamma Z_i, \label{equ:iv10}\\
Z_i & = \sum_{j=1}^M z_{i,j}, \label{equ:iv20}\\
\dot{z}_{i,j} & = - a_1 z_{i,j+1} (z_{i,j+2} -z_{i,j-1})
-a_2 z_{i,j}+X_i, \label{equ:iv30}
\end{align}
and the corresponding observations
\beqn{equ:iv40}
\eta_k = \sum_i c_{ki} (X_i + r_i), 
\eeq
where $a_1=100$, $a_2=10$, $X_i$ and $z_{i,j}$ denote the slow and
the fast degrees of freedom respectively, $F=18$, $\gamma$ is a
constant quantifying the influence of the fast degrees of freedom onto the slow ones, and $r$ is short time correlated Gaussian noise with $\mathbb{E}[r_i(t)r_j(t)] = \rho_0 \delta_{ij}$.
Two cases of $C = \{c_{ij}\}$ are considered.
In the first case, observations are available from all degrees of freedom, that is, $c_{ij} = \delta_{i,j}$.
In the second case, only every second degree of freedom was observed, that
is, $c_{ij} = \delta_{2i,j}$.
The index of the slow degrees of freedom as well as the first index of the fast
ones takes the values $i=1,\ldots, L$, periodic boundary conditions being
assumed.
The second index of the fast degrees of freedom takes the values $j=1,\ldots,M$,
where $z_{i, M+1} = z_{i+1, 1}$, $z_{i, M+2} = z_{i+1, 2}$, and $z_{i, 0} = z_{i-1, M}$, for $i = 1 \ldots
L$.
In all the examples considered here, $L=64$ and $M=8$.
In any event, only slow degrees of freedom are observed.
Simulations of the reality have been generated by integrating
Equations~(\ref{equ:iv10}--\ref{equ:iv30}) by means of a simple Euler scheme
with integration step $\delta t = 10^{-5}$.
Observations $\eta_i(t)$ were sampled with $\Delta t=5.12 \cdot 10^{-3}$ only.
The corresponding sampling frequency is still sufficiently large, compared to the bandwidth of the signal.
As the assimilating model we use the Lorenz'96 system with one scale only (i.e.\ no fast degrees of freedom)
\begin{align}
\dot{x}_i & = -x_{i-1} (x_{i-2} -x_{i+1})
-x_{i}+ F, \labeln{equ:iv50}\\
y_i & = \sum_i c_{ki} x_i \labeln{equ:iv60}.
\end{align}
This setup is motivated by practical situations in which typically high frequency modes living on small scales cannot
be taken into account, due to limited computational power and
impossibility of acquiring data at such small scales.
Note that in the example presented here, we have chosen the same forcing term
$F$ for both model and reality, whence any model error is due to the absence of
the fast degrees of freedom in the model.
Thus, $\gamma$ in Equation~\eqref{equ:iv10} controls the amount of model error.
In the following, we consider four different scenarios.
The first two are given by a small noise case with $\rho_0 = 0.01$
(corresponding to a SNR of~57dB) and $\gamma=0.01$ as well as a large noise
case with $\rho_0 = 1$ (corresponding to a SNR of~16dB) and $\gamma=1$.
Observations were taken of all slow degrees of freedom, corresponding to $c_{ij} = \delta_{i,j}$, as discussed above.
For the other two scenarios, the same combinations of the
noise values are considered, but observations are only taken of
every second slow degree of freedom, that
is, $c_{ij} = \delta_{2i,j}$.
\subsection{Variational data assimilation}
\label{subsec:numerical-experimentsVDA}
The functional to be minimised (\ref{equ:30}) takes the form
\begin{multline}
\label{equ:iv70}
\overline{A}_\alpha = \int_0^T \! \dd t \;
\frac{\alpha}{2} \sum_k \Bigl(\eta_k - \sum_i c_{ki} x_i \Bigr)^2 +         \frac{1-\alpha}{2} \sum_i u_i^2\\
+ \sum_i \lambda_i \bigl( \dot{x}_i + x_{i-1} ( x_{i-2} - x_{i+1}) + x_{i}- F - u_i \bigr),
\end{multline}
where the total integration time was $T = 2^{20}\delta t$, the $u_i$ denote
the dynamical perturbation, and $\lambda_i$ are Lagrange multipliers.
In the present numerical example, the matrices $R$ and $Q$ were taken as unit matrices.
This is justified since the variability of the different dynamical degrees of freedom are expected to be similar, and likewise for the output.
After eliminating $u$ with the help of $u_i=-\lambda_i/(1-\alpha)$, the Hamiltonian equations resulting from the functional $A$ are given by
\begin{align}
\dot{x}_i & = -x_{i-1} (x_{i-2} -x_{i+1})
-x_{i}+ F-\frac{\lambda_i}{1-\alpha},\\
\dot{\lambda}_i
& = \lambda_{i+1}(x_{i-1}-x_{i+2}) + \lambda_{i+2} x_{i+1} - \lambda_{i-1} x_{i-2} + \lambda_i \nonumber \\
& \quad + \alpha \sum_k c_{ki} ( \eta_k - \sum_j c_{kj} x_j).
\end{align}
These equations have been solved by means of a NAG~boundary value problem solver \citep[D02RAF, see for example][]{pereyra78} with boundary values
$\lambda(0)=\lambda(T)=0$.
The resolution $\delta t_\mathrm{model}$ of the BVP~solver mesh is in general not identical to the sampling interval $\Delta t$ of the observations.
For intermediate times at which no observations exist, $\eta_i(t)$ was
interpolated by means of cubic splines.
In order to determine the sensitivity, we proceed according
to Section~\ref{subsec:sensitivity-vda-bvp}, that is, first $P$ is
obtained by means of the matrix valued equation~(\ref{equ:210.4}),
which is integrated backward in time, and subsequently Equation~\eqref{equ:260} is integrated forwards in time to get $J$.
In the present situation, the matrices
$F$, $M$, and $N$ are given by
\begin{align}
F_{i,j} & = -\delta_{i,j} + \delta_{i,j+1}(x_{i+1}-x_{i-2})\nonumber \\
 & \quad + (\delta_{i,j-1}-\delta_{i,j+2})x_{i-1}, \\
M_{i,j} & = -\frac{\delta_{i,j}}{1-\alpha},\\
N_{i,j} & = \alpha \sum_k c_{ki} c_{kj} - \delta_{i,j+1} \lambda_{i+1} +
\delta_{i,j+2}\lambda_{i-1} \nonumber \\
 & \quad - \delta_{i,j-1}\lambda_{i+2} + \delta_{i,j-2}\lambda_{i+1},
\end{align}
respectively.
The sensitivity, $S$, is eventually obtained according to Equation~\eqref{equ:262}, or precisely
\begin{equation}
S = -\alpha \Delta t \int_0^T \! C J_t C^T \dd t,
\end{equation}
where $\Delta t$ is the sampling interval of the observations.
Note again that $\Delta t$, in general, need not coincide with either the resolution $\delta
t_{\mathrm{model}} = 1.28 \cdot 10^{-3}$ at which Equations~(\ref{equ:210.4}) and~(\ref{equ:260}) were integrated, nor with the resolution of the the BVP solver's time mesh.
\begin{figure*}
\centering
\includegraphics[width=0.8\textwidth]{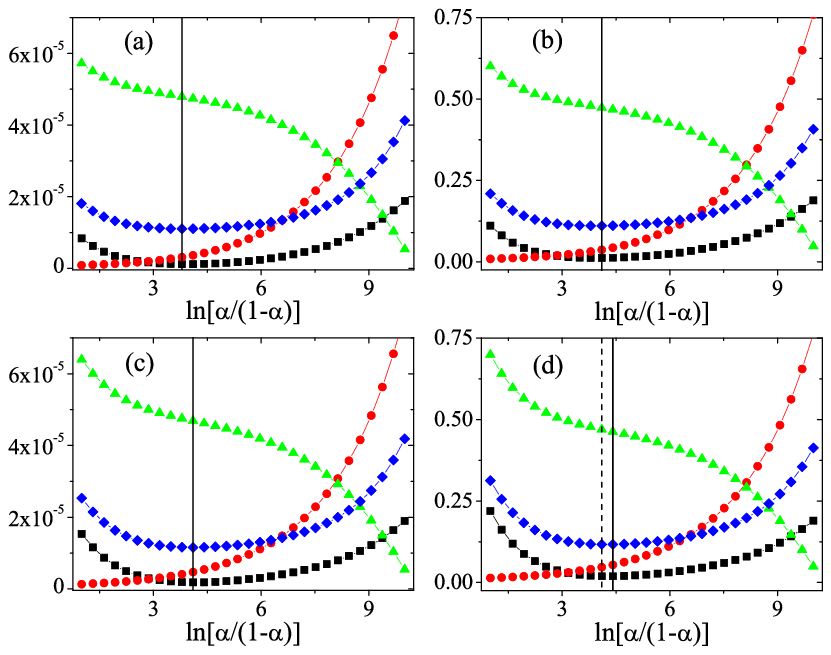}
\caption{\label{fig1}
For variational assimilation, $E_\mathrm{oos}$ ($\blacklozenge$),
$A_\mathrm{T}$ ($\blacktriangle$), sensitivity ($\bullet$), and
$A_\mathrm{A}$ ($\blacksquare$) are plotted
vs.~$\ln{[\alpha/(1-\alpha)]}$, for the small noise complete
information (a), large noise complete information (b), small noise
incomplete information (c), and large noise incomplete information
(d) cases.
The solid vertical line marks the value of the coupling constant
at which the assimilation error gets minimal,
$\alpha_\mathrm{opt}$. The position of the minimum of
$E_\mathrm{oos}$ coincide perfectly with $\alpha_\mathrm{opt}$ in
all cases but (d), where the minimum of $E_\mathrm{oos}$ is marked
by the dashed vertical line. Note that, even in the latter case,
the deviation of the two minima is small. $E_\mathrm{oos}$ and
$A_\mathrm{T}$ have been shifted on the ordinate for better
visibility.}
\end{figure*}
Figure~\ref{fig1} displays $\eloo$ 
(diamonds), computed according to Equation~\eqref{equ:150}, approximating the average (diagnostic) tracking error as in~\eqref{equ:90}.
For this study, we set $W = R = $unity~matrix.
Also shown is the sensitivity (more specifically, $2 \tr(W S \rho_0)$, bullets), the tracking error $A_\mathrm{T}$
(triangles) and the assimilation error $A_\mathrm{A}$ (squares)
for the four studied cases (small noise and complete information on panel~(a);
large noise and complete information on panel~(b), small noise and incomplete
information on panel~(c), large noise and incomplete information on panel~(d)).
All quantities are shown versus the sensitivity parameter $\alpha$.
As a reminder, the assimilation error $A_\mathrm{A}$ is given by the first term on the right hand side in Equation~\eqref{equ:84}.
In Figure~\ref{fig1}, $A_\mathrm{T}$ and $\eloo$ have been shifted on the
ordinate for better comparison.
Furthermore, all quantities shown in this section have been divided by the time span $T$, and the number of observed degrees of freedom $d$.
In other words, all quantities should be interpreted per unit time, and per observed degree of freedom.
The tracking error as well as
assimilation error have been integrated using the trapezoidal rule with resolution given by $\Delta t$.
As expected, $A_\mathrm{T}$
decreases monotonously with $\alpha$, while the opposite is the
case for the sensitivity.
This alone does not imply a minimum of $\eloo$, for which a cancellation of the derivatives of $A_\mathrm{T}$ and $2 \tr(W S \rho_0)$ with respect to $\alpha$ is necessary. 
Anyway, $\eloo$ displays a well defined minimum at a value of the sensitivity
parameter $\alpha$ which coincides almost perfectly with the value for $\alpha$
with minimum assimilation error.
In Figure~\ref{fig1}, the positions of the
minima of the assimilation error and the linearised out--of--sample
error are marked by the solid and dashed vertical lines,
respectively.
In almost all cases, the minima of $\eloo$ and the assimilation error coincide
perfectly within the resolution by which the sensitivity parameter $\alpha$
has been sampled.
Only for the case of large noise and observations of every second degree of freedom only, a discrepancy is observed, which however is still very small.
The dependence of the optimal $\alpha$ (in terms of a minimal tracking error) on the experimental setup, in particular on the amplitudes of both the dynamical and observational perturbations, was already investigated in~\citet{broecker_variational_2010}.
In that study, it emerged that the minimum of the assimilation error moves
towards smaller values of $\alpha$ with increasing observational noise, and
towards larger values with increasing dynamical perturbations.
This effect is encountered in the present experiments as well. 
Bearing in mind that the sensitivity increases with increasing $\alpha$, this means qualitatively that the larger the observational noise, the smaller the sensitivity should be; larger dynamical perturbations though require a larger sensitivity.
According to Equation~\eqref{equ:84}, the difference between the assimilation error $A_{\mathrm{A}}$ and the out--of--sample error $\eloo$ should be independent of $\alpha$, and be equal to $(t_f - t_s)\tr (W \rho_0)$.
In our numerical results, this is not quite the case for very large $\alpha$ (i.e.\ much larger than the optimal value).
A possible explanation is that the sensitivity was estimated using the linear approximation~\eqref{equ:180}, which essentially assumes a small response of solutions to changes in the input.
Clearly, this is not quite true for large $\alpha$.
Away from extremely large values of $\alpha$ though, we found the difference between the assimilation error $A_{\mathrm{A}}$ and the out--of--sample error $\eloo$ not only to be independent of $\alpha$ (as can be discerned already from Fig.~\ref{fig1}), but also quantitatively in accordance with what our theory predicts.
We can conclude that, despite several approximations, our approach yields quantitatively correct estimates of the out--of--sample error for relevant ranges of the sensitivity parameter~$\alpha$.
\begin{figure}
\centering
\includegraphics[width=\columnwidth]{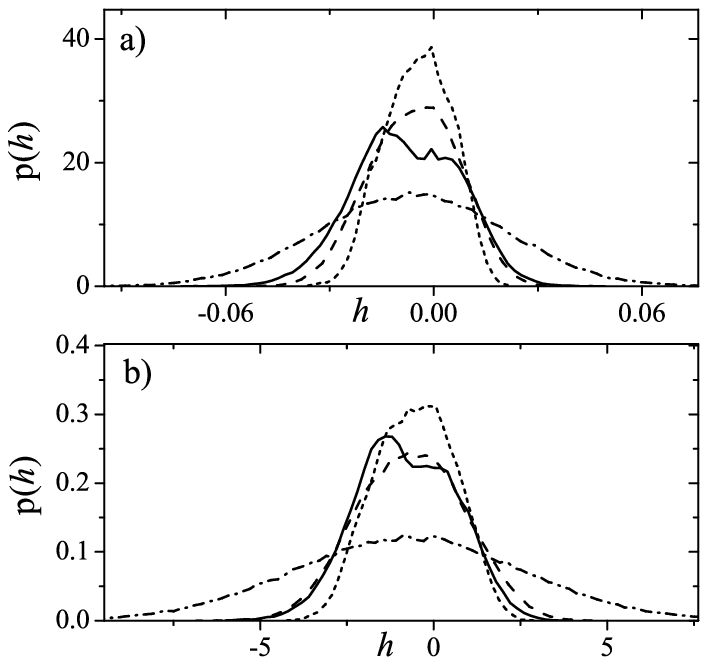}\\[1ex]
\begin{tabular}{cccc}
	& $\alpha_1$ & $\alpha_2$ & $\alpha_3$ \\
panel (a) &	2.55 & 4.1  & 5.66 \\  
panel (b) & 2.86 & 4.41 & 5.97
\end{tabular}
\caption{\label{fig11}
The distributions of the actual fast degrees of freedom (more specifically $-\gamma Z_i$ and and the dynamical perturbations $u_i$ are shown for
three values of $\alpha$.
The variable $h$ on the ordinate stands for either $-\gamma Z_i$ and $u_i$.
Panel~(a) shows the small noise case, Panel~(b) the large noise case.
Both cases used complete information about the slow degrees of freedom.
In both panels the solid line shows the distribution of $Z_i$, while the dotted, dashed, and dash--doted lines show the distribution of $u_i$ for three different (increasing) values of $\alpha$.
In both panels, the dashed line corresponds to $\alpha$ with a minimal out--of--sample error.
The table gives the actual values of $\log(\frac{\alpha}{1 - \alpha})$ corresponding to the displayed cases.
}
\end{figure}
Another question which might arise naturally is to what degree the dynamical perturbations $u$ carry information about the unmodelled degrees of freedom. 
We might expect, provided ``all went well'', that the perturbations are similar to the unmodelled degrees of freedom at least in some statistical sense.
Learning something about the true underlying dynamics is a possible application of the concepts proposed in this paper, which will be subject to further investigation. 
A few very preliminary results shall be presented here.
In Figure~\ref{fig11}, several distributions (in the form of probability density functions) of the dynamical perturbation $u_i$ are shown.
Panel~(a) and~(b) correspond the small noise case and the large noise case,
respectively; both cases used complete information on the slow degrees of
freedom.
Both panels shows distributions for $u$ corresponding to three different values of $\alpha$ (dashed, dotted, and dash--dotted lines, in increasing order of $\alpha$).
In both panels, the dashed line corresponds to that $\alpha$ which gave a minimal out--of--sample error.
Further to that, the distribution of the sum over the fast, unmodeled degrees of
freedom is shown, more specifically, the distribution of $-\gamma Z_i$ (solid line).
Pending a more detailed analysis, visual inspection already shows, reassuringly, that the
distribution of the unmodeled degrees of freedom agrees best with the
distribution of $u$ for the optimal value for $\alpha$, that is, the $\alpha$ value giving a minimal out--of--sample error.
For $\alpha$~values smaller than the optimal one, the distribution is too wide, while a too large value of $\alpha$ gives a distribution of perturbations $u$ which appears to be narrower than that of the fast degrees of freedom.
The actual values for $\alpha$ (or rather, for $\log(\frac{\alpha}{1 - \alpha})$ for comparison with Fig.~\ref{fig1}) can be found in the table complementing Figure~\ref{fig11}.
% 
% were $\alpha_1 \simeq .92$,
% $\alpha_2 \simeq .98$, $\alpha_3 \simeq .996$ for panel~(a),
% and $\alpha_1 \simeq .94$, $\alpha_2 \simeq .988$, $\alpha_3 \simeq .997$ for panel~(b).
%
% \begin{table}
% \begin{center}
% \end{center}
% \caption{\label{tab:alphas}}
% \end{table} 
% 
%

%
It is worth stressing that the the estimated distributions (or other statistical
properties) of the perturbations will depend not only on the actual model error but also on the specific assimilation scheme.
In the present case for example, the fast degrees of freedom seem to have a nonzero mean value, while the distributions of $u$ are centered closer to zero.
This is presumably due to the specific form of the functional $A$ (Equ.~\eqref{equ:iv70}), which clearly favours a perturbation $u$ with a small mean value. 
In other words, as estimators of the fast degrees of freedom, the $u_t$ are expected to be biased towards zero.
This may be remedied by introducing an offset in
the penalization term for the control in the functional $A$.
This offset can than be treated as an additional control parameter and its
optimal value may again be estimated by the method introduced in this article.
Concerning the equivalence of the two approaches for calculating the sensitivity
in variational data assimilation problems (as detailed in
Subsections~\ref{subsec:sensitivity-vda-colloc-jac}
and~\ref{subsec:sensitivity-vda-bvp}), numerical experiments were carried out
employing the Lorenz'63 system~\citep{lorenz63}.
These experiments will not be discussed in detail here;
For a comprehensive report of the experiments, see~\citet{broecker_variational_2010}.
The Lorenz'63 system is a simple dynamical system with three degrees of freedom, which exhibits chaotic motion.
For systems of this size, the collocation approach leads to perfectly manageable problems.
The sensitivity was calculated using the methodology of both Subsection~\ref{subsec:sensitivity-vda-colloc-jac} and~\ref{subsec:sensitivity-vda-bvp}.
The results turned out to be in perfect agreement.
\subsection{Statistical interpretation of the optimal $\alpha$}
\label{subsec:numerical-experiments-alpha-interp}
It was already mentioned that there is an alternative interpretation of the variational approach, namely as a BLUE or more generally a maximum--aposteriori estimator. 
One of the problems with this interpretation was that it required the weighting matrices $\alpha R$ and  $(1 - \alpha) Q$ to be equal to the inverse observational and dynamical error covariances, respectively. 
Given that we now have a methodology to find $\alpha$, the question arises whether this provides us with reasonable estimates of observational and dynamical error covariances.
In order to investigate this question further, we carried out several simulations, with the following setup.
We again generated ``reality'' using the Lorenz'96 equations~\eqref{equ:iv10}, but replaced the term $-\gamma Z_i(t)$ by white noise $g_i(t)$ with covariance function $\E{g_i(t)g_j(s)} = q \delta_{ij} \delta(t_1 - t_2)$.
Equation~\eqref{equ:iv10} (with $L = 32$ degrees of freedom) was then integrated using a stochastic Euler scheme.
The observations (all degrees of freedom were observed) where created as before, and the data assimilation machinery was applied, including determining the out--of--sample error and the optimal $\alpha$.
Ideally, we would like to compare $\alpha$ with a version of BLUE in continuous time and with $\delta$--correlated disturbances.
Pending a more rigorous discussion of such a theory, we have to make do here with the following heuristics. 
As was discussed in Section~\ref{subsec:noise-assumptions}, we assume the observational noise process $r_{t}$ do be interpolated from samples of a process $\nu_t$ with short correlation and variance $\rho_0$.
Since the integral over the entire power spectrum of $\nu_t$ must be equal to $\rho_0$, it is, strictly speaking, impossible that $\nu_t$ has a flat power spectrum with truly unlimited bandwidth. 
However, we might alternatively obtain $r_t$ by {\em low--pass filtering}  (with appropriate cutoff) a white noise process $\nu_t$ with correlation function $\E{\nu_t \nu_s} = \sigma \cdot \delta(t - s)$; in order that $r_t$ has the power spectrum required in Equation~\eqref{equ:400}, we need to set $\sigma^2 = \rho_0 \cdot \Delta t$. 
We tentatively interprete the action integral~\eqref{equ:20} as a finite bandwidth approximation of the logarithmic aposteriori of the orbit $\{x_t, t \in [t_s, t_f]\}$, for infinite bandwith observational and dynamical noise.
(At present, we believe that the action integral~\eqref{equ:20} needs to be modified in order to survive the limit of infinite bandwidth.)
Given this interpretation is correct, we should have 
\beq{equ:iv80}
\log \frac{\alpha}{1 - \alpha} = \log{\frac{q}{\sigma}} 
= \log{\frac{q}{\rho_0 \cdot \Delta t}}.
\eeq
Our numerical experiments indicate that relation~\eqref{equ:iv80} is indeed correct.
In total, 15~simulations were run, with both $\rho_0$ and $q$ ranging between $0.01$ and $2$.
The simulations where collated in five groups of three each, with $q/r$ constant within each group; the five groups corresponed to the ratios $q/r = [1.5, 1, 2, 5, 10]$.
We then determined $\alpha_s$ by optimising the out--of--sample error. 
Further, $\alpha_{\text{opt}}$ optimal for the true assimilation error was recorded.
\begin{figure*}
\centering
\includegraphics[width=\textwidth]{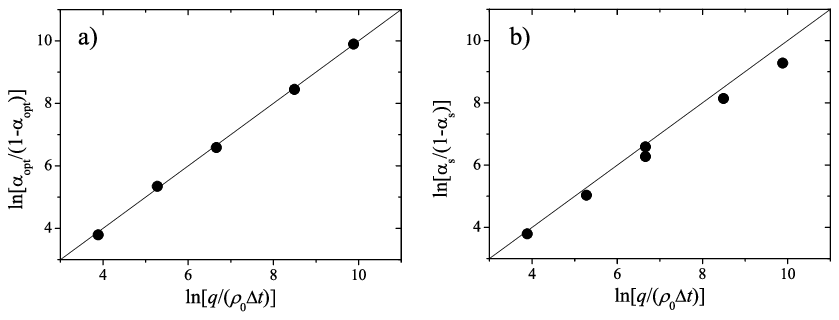}\\[1ex]
\caption{\label{fig:blue-test}
$\log \frac{\alpha}{1-\alpha}$ is plotted vs.\ $\log \frac{q}{\rho_0\Delta t}$, with
$\alpha$ being the optimal value for the assimilation
error in plot~(a), resp.\ the out of sample error in plot~(b). Various choices of $q$ and
$\rho_0$ are shown. The straight line has been included to guide the eye and corresponds to
$\log \frac{\alpha}{1-\alpha}=\log \frac{q}{\rho_0\Delta t}$.}
\end{figure*}
In Figure~\ref{fig:blue-test}, plot~(a), $\log \frac{\alpha_{\text{opt}}}{1 - \alpha_{\text{opt}}}$ is shown versus $\log{\frac{q}{\rho_0 \cdot \Delta t}}$.
Clearly, the two values agree almost perfectly.
In Figure~\ref{fig:blue-test}, plot~(b), $\log \frac{\alpha_s}{1 - \alpha_s}$ is shown versus $\log{\frac{q}{\rho_0 \cdot \Delta t}}$, again with very good agreement. 
A few further simulations were run with different $\Delta t$ (not shown), confirming that the scaling with $\Delta t$ is indeed as indicated by Equation~\eqref{equ:iv80}.
The conclusions we can draw from these investigations are that firstly, since $\alpha_s$ is very close to $\alpha_{\text{opt}}$, our method again provides an $\alpha$ which very nearly minimises the assimilation error. 
Secondly, we see that $\alpha_{\text{opt}}$ is indeed very close to the logarithmic ratio of the noise intensities; thereby, the presented methodology might also be seen as a method for estimating the dynamical noise intensity. 
These results clearly call for further theoretical investigation, which has to be deferred to a future paper.
From the heuristics presented above, it seems that the variational approach to data assimilation as studied here can indeed be regarded as a finite bandwidth approximation to a maximum--aposteriori trajectory estimator for problems with white observational and dynamical noise.
As mentioned in Section~\ref{sec:variational-data-assimilation}, \citet{zeitouni_maximum_1987} discuss a maximum--aposteriori concept, but their expression for the logarithmic aposteriori differs from our action integral by a term involving the divergence of the vector field $f$ (and other terms that do not bear on the minimisation).
Interestingly though, the divergence of the Lorenz'96 system is constant; therefore in the present situation our action integral is equivalent to the maximum--aposteriori concept of Zeitouni and Dembo, if relation~\eqref{equ:iv80} is in force.
We speculate that in general, relation~\eqref{equ:iv80} is true only if the correct form of the logarithmic aposteriori is employed. 
Encouraged by these findings, we might look back at Section~\ref{subsec:numerical-experimentsVDA} and see if an effective white noise can be associated with the model error term $- \gamma Z_i$ despite the fact that there, the $Z_i$ were not stochastic but the fast degrees of freedom. 
From our numerical simulations, we estimated $\mbox{Var}[\gamma Z_i]$ to be about $ 2.28 \cdot 10^{-4}$ and $2.016$ for the small resp large noise cases considered in Section~\ref{subsec:numerical-experimentsVDA} (cf.\ also Fig.~\ref{fig11}).
For the ratio $ \frac{\mbox{Var}[\gamma Z_i]}{\rho_0}$ we obtain $0.4453$ and     $0.3937$; the corresponding values for $\frac{\alpha}{1 - \alpha}$ were $60.3403$ and $82.2695$.
Unfortunately, we were unable to relate the noise ratio with the corresponding $\alpha$--values.
Note that the variance ratio is not proportional to the $\alpha$--ratio, so a simple rescaling with some effective sampling time, for example, could give a very approximate correspondence at best. 
In fact, simply scaling with $\delta t$ yielded completely wrong results.
We conclude that for the purpose of variational data assimilation, interpreting the fast degrees of freedom as white noise model error can be very dangerous indeed.
Clearly, the fast degrees of freedom differ from white noise in a large number of ways, but it would still be interesting to know why we see so different behaviour between these experiments and those with white noise perturbances.
This will be subject to future research. 
\subsection{Synchronisation}
\label{subsec:numerical-experimentssynch}
As a second example, we have studied assimilations by means of
synchronisation.
In the particular setup we studied, the model is
coupled to the observations through a simple linear coupling term
\beqn{synchronumerics}
\dot{x}_i = -x_{i-1} (x_{i-2} -x_{i+1})
-x_{i}+ F + \kappa (\eta_i-x_i),
\eeq
where $\kappa$ is the coupling constant.
We can expect that, if
the coupling is too strong, the observations will be tracked too
rigorously and hence observational noise is not filtered out
well.
On the other hand, if $\kappa$ takes to small values, the
observations are tracked poorly and, as an additional
consequence, model errors are not compensated for.
Hence, again, we can expect the assimilation error to take a minimum at some
nontrivial value of $\kappa$.
For the numerics, the setup described in Section~\ref{sec:theory-synchronisation} with $c_{i,j} = \delta_{i,j}$ for a total time interval of length $T=2^{22} \cdot \delta t$ was simulated, where $\delta t = 10^{-5}$ and observations where
taken at sampling intervals $\Delta t = 5.12 \cdot 10^{-3}$.
The model was
integrated with a time step $\delta t_\mathrm{model}=2^{4}\delta
t=1.6 \cdot 10^{-4}$.
Again, the observations $\eta_i(t)$ where
interpolated by means of cubic splines at
intermediate times where no observations where recorded.
For the simple synchronisation setup studied here, we get from Equation~\eqref{equ:440} that the sensitivity
per unit time and per observed degree of freedom is given by $\frac{1}{2} \kappa \Delta t$.
To calculate the tracking and assimilation errors,
a transient time was ignored to give the system
time to synchronise.
\begin{figure*}
\centering
\includegraphics[width=0.8\textwidth]{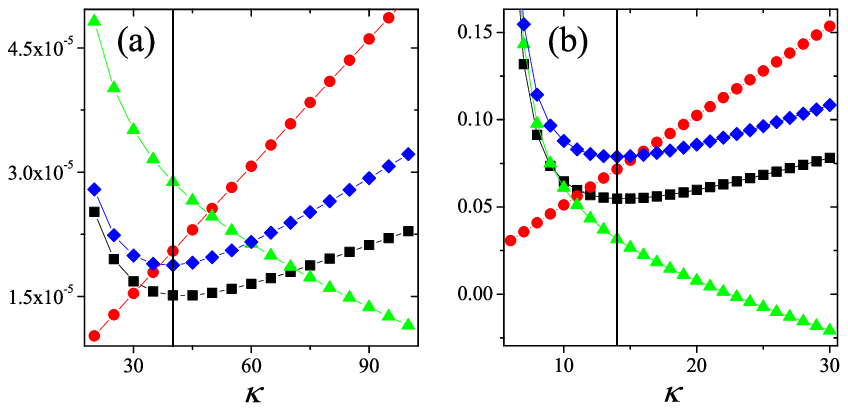}
\caption{For synchronisation, $E_\mathrm{oos}$ ($\blacklozenge$),
$A_\mathrm{T}$ ($\blacktriangle$), $2 \sigma^2 S$ ($\bullet$), and $A_\mathrm{A}$
($\blacksquare$) are plotted vs. $\kappa$, for the small noise (a) and
large noise (b) cases. Observations where taken of all slow
degrees of freedom. The vertical line marks the value of the
coupling constant at which the assimilation error gets minimal,
$\kappa_\mathrm{opt}$. Note that the position of the minimum of
$E_\mathrm{oos}$ coincides perfectly with $\kappa_\mathrm{opt}$.
$E_\mathrm{oos}$ and $A_\mathrm{T}$ have been shifted on the ordinate for better
visibility.}\label{fig2}
\end{figure*}
In Figure~\ref{fig2} the out--of--sample error
(diamonds) is presented, together with the tracking error (triangles), the
sensitivity as $2 \rho^2 S$~(bullets), and the assimilation error (squares) for
the synchronisation scenarios corresponding to the two cases (weak
noise (a) and large noise (b)) with complete information of the
slow degrees of freedom, for various choices of the coupling
parameter $\kappa$.
Again, $A_\mathrm{T}$ and $\eloo$
have been shifted on the ordinate for better comparison.
Just as in the case of
the variational assimilation, the tracking error decreases
monotonously with increasing coupling strength, while the
sensitivity increases monotonously.
Again, the linearised out--of--sample error shows a well defined minimum at a certain value of
$\kappa_\mathrm{opt}$, which coincides perfectly, within the
studied resolution, with the $\kappa$ at which the assimilation
error is minimal.
To guide the eyes, we plot a vertical line to
mark the positions of the minima.
The extremely high values of
$\kappa_\mathrm{opt}$ in the low noise case can be easily
understood when having in mind that, due to the the small
observational noise, a rigorous tracking of the observed signal does
not introduce large dynamical perturbations.
\begin{figure}
\centering
\includegraphics[width = \columnwidth]{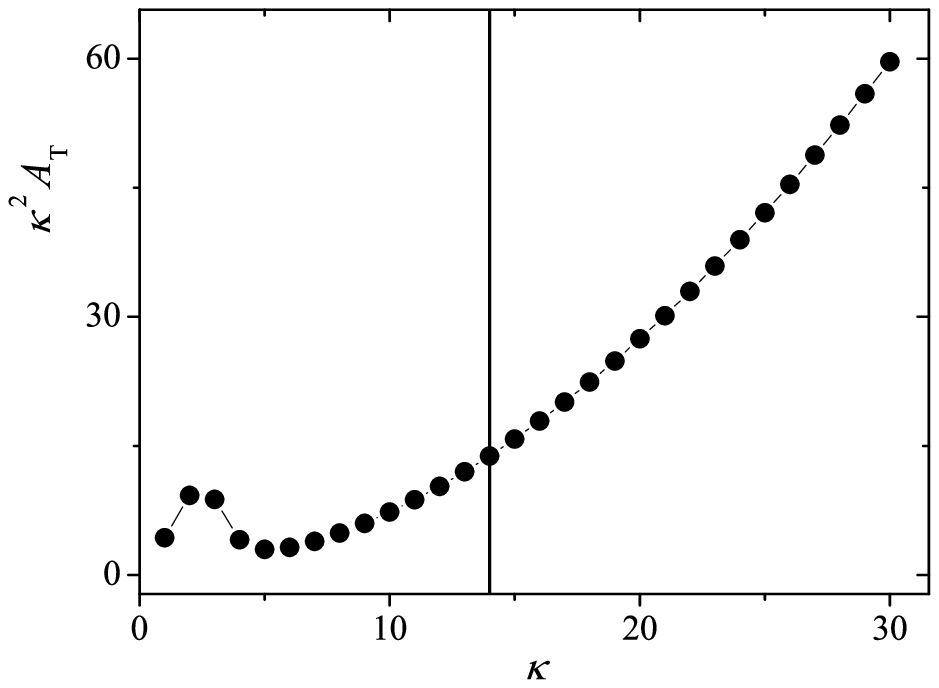}
\caption{The squared coupling term, $\kappa^2A_\mathrm{T}$, is
plotted vs. $\kappa$ ($\bullet$). The vertical line denotes the
value of the coupling strength at which the assimilation error
gets minimal, $\kappa_\mathrm{opt}$. Note that the minimum of
$\kappa^2A_\mathrm{T}$ does not coincide with
$\kappa_\mathrm{opt}$.}\label{fig3}
\end{figure}
As an interesting aside, note that although the
tracking error decreases monotonically when increasing $\kappa$,
the strength of the dynamical perturbations,
$\int_{t_s}^{t_f}\mathrm{d}t\sum_{i=1}^L[\kappa(\eta_i-x_i)]^2/[(t_f - t_s)L]=\kappa^2
A_\mathrm{T}$, does not decrease.
In fact, it displays a well defined
minimum, meaning that, at this value of $\kappa$, the influence of
the coupling term gets minimal.
Note that this value of $\kappa$ does not correspond to the $\kappa$ value at
which the out--of--sample error gets minimal.
To demonstrate
this statement, we plot $\kappa^2 A_\mathrm{T}$, for the large
noise case with observations at all degrees of freedom (see Fig.~\ref{fig3}).
The vertical line marks $\kappa_\mathrm{opt}$, at
which the assimilation error and the linearised out--of--sample
error display their minima.
Note that the minimum of the coupling
term lies far away from $\kappa_\mathrm{opt}$.
Rather, it appears
that the minimum of $\kappa^2 A_\mathrm{T}$ is determined by the
value of $\kappa$ at which, in the perfect model case, the phase
transition would be expected.

The coupling term, in case of synchronisation, is in some sense
analogous to $u_t$ in the variational assimilation case, whence
$\kappa^2 A_\mathrm{T}$ is analogous to $A_\mathrm{M}$ in that it
describes the average deviation from the pure model.
In contrast
to $\kappa^2 A_\mathrm{T}$ though, $A_\mathrm{M}$ is monotonously
increasing with increasing sensitivity.
On the other hand,
it might be argued that the analogy between $\kappa^2
A_\mathrm{T}$ and $A_\mathrm{M}$ is warranted only far beyond the
phase transition.
In that range, $\kappa^2 A_\mathrm{T}$ and
$A_\mathrm{M}$ show qualitatively the same behaviour.
%
%
%%%%%%%%%%%%%%%%%%%%%%%%%%%%%%%%%%%%%%%%%%%%%%%%%%%
%
\section{Conclusions and outlook}
When assimilating observational data into a dynamical model, then
solutions which closely track the observations typically do not
exactly adhere to the model equations, while enforcing the latter
would cause unacceptably large deviations from the observations.
Thus, in data assimilation one faces a fundamental trade--off,
caused by model error.
This trade--off is investigated in this paper.
To settle the trade--off in real--world situations, a minimal out--of--sample error is proposed as a criterion.
As was shown, the out--of--sample error is connected to the
assimilation error, but can also be expressed by
means of the tracking error and the sensitivity of the
assimilation scheme.
The sensitivity quantifies by how much the output of the data
assimilation depends on perturbations of the observed time series.
The tracking error is available under operational circumstances.
It is demonstrated that also the sensitivity can, at least in principle, be estimated operationally.
The details depend on the specifics of the employed data
assimilation algorithm.
In this paper, both variational data assimilation as well as synchronisation (aka nudging) are
looked at.
It is stressed that when calculating the sensitivity, several
quantities can (and should) be recycled which are available
already from the data assimilation algorithm proper, thereby
saving computational resources.
The feasibility of the proposed schemes is demonstrated through
numerical experiments involving the Lorenz'96 system.
The variational approach studied here might be interpreted as a (continuous time nonlinear) analogy of the best linear unbiased estimator (BLUE).
This interpretation comes with a specific recommendation as to how to set the sensitivity parameter, namely as the ratio between the dynamical and observational error covariances.
This recommendation was tested against the method investigated here;
we found that if the assumptions behind BLUE apply, both methods gave essentially the same results (yet ours needs less information to be applicable).
For model error consisting of fast dynamical variables though, we found BLUE to perform very poorly.
In the future, it would be desirable to apply the presented method
to realistic weather models.
Such models typically include various inequivalent degrees of
freedom, necessitating more than just one sensitivity parameter.
In principle, the approach is straightforwardly applicable to the
case of multiple sensitivity parameters, although exploring a
larger dimensional parameter space will of course increase the
numerical costs considerably.
In order to keep things concise, simple quadratic forms for the
penalisation terms are used in this article.
Furthermore, simple additive observational errors and dynamical
perturbations are chosen.
In realistic situations, more involved choices might be
appropriate.
However, for the most part, the discussion does not rely on these choices, but readily applies to more complicated cases.
In any event, the a priori knowledge of the observational noise
characteristics is crucial for good parameter estimations.
It would therefore be of interest to study in more detail the
impact of wrong assumptions concerning the observational  noise on
the obtained solutions.
Furthermore, in more realistic setups, the observational noise might
display correlations, in which case the methodology still applies,
but requires more involved calculations.
The proposed procedure
should not only be applicable to determining optimal sensitivity
parameters but also to obtain better estimations of model
parameters.
Again, the advantage of the proposed procedure would be that parameters are
assessed by means not of the (in sample) tracking error but rather of the
assimilation error, which is a better performance measure for data assimilation
systems.
Finally, a general discussion as to which measure for assimilation performance is appropriate to determine the sensitivity parameter would be desirable.
While the minimisation of the assimilation error appears to be a
sensible choice for the assessment of trajectories obtained by
assimilation, there have been other suggestions in the past.
For example, in~\citet{szendro_problem_2009} it has been proposed
to assess the quality of an assimilated solution taking into account the
spatial structure of the assimilation error.
The reason is the observation that even if the assimilation error is small, the reconstructed trajectory does not necessarily correspond to a typical trajectory of the reality, and might therefore yield poor predictions.
Nonetheless, numerical experiments (not shown here) indicate that
sensitivity parameters optimal with respect to the assimilation error do not coincide with sensitivity parameters according to the methodology proposed in~\citet{szendro_problem_2009}.
%
%
%%%%%%%%%%%%%%%%%%%%%%%%%%%%%%%%%%%%%%%%%%%%%
%
%
\section*{Acknowledgements}
Fruitful discussions with Kevin Judd, Markus Niemann, and Holger Kantz are
gratefully acknowledged.
Further, we thank two anonymous referees for valuable suggestions, and for insisting on a comparison with BLUE.
%
%
%%%%%%%%%%%%%%%%%%%%%%%%%%%%%%%%%%%%%%%%%%%%%
%
\section*{Appendix: Structure of the collocation equations\\ for the BVP solver \texttt{bvp4c}}
%
%
% \label{apx:bvp4c}
%
In this appendix, we will be assuming that the BVP is given as
\beqn{equ:a265}
\dot{z}_t = F(t, z_t), \qquad t \in [t_s, t_f], \qquad b(z_{t_s}, z_{t_f}) = 0,
\eeq
with $z_t \in \R^{2D}$, $F$ a vector field on $\R^{2D}$ and $b(., ..)$ a function representing the boundary conditions.
Suppose $t_s = t_0 < \ldots < t_N = t_f$ are temporal mesh points; write $z_i := z_{t_i}$ and $\z := (z_0, \ldots, z_N)$ for the values of the solution at these points.
The collocation equations
\beqn{equ:a270}
\Phi_i(\z) = 0, \qquad i = 0 \ldots N,
\eeq
are solved to obtain $\z$.
The BVP solver \texttt{bvp4c}~by \citet{kierzenka01} implemented in Matlab uses a fourth order implicit Runge--Kutta scheme to approximate the differential relation, in which case
\beqn{equ:280}
\Phi_i(\z) = z_i - z_{i-1} - \frac{h_{i-1}}{6}
    \left( F_{i-1} + 4 F_{i-1/2} + F_{i}
    \right)
\eeq
for $i = 1 \ldots N$ with $h_i = t_{i+1} - t_{i}$, $t_{i-1/2} = t_{i-1} + \frac{h_{i-1}}{2}$, and
\begin{align*}
F_i & = F(t_i, z_i), \\
F_{i-1/2} & = F \left( t_{i-1/2},
    \frac{z_{i-1} + z_{i}}{2} - \frac{h_{i-1}}{8} \left( F_{i} - F_{i-1} \right)\right).
\end{align*}
Finally, $\Phi_0(\z) = b(z_0, z_N) $ represents the boundary conditions.
Using these expressions, we get the following partial derivatives of $\Phi$:
\beqn{equ:340}
\frac{\partial \Phi_0}{\partial F_{j}} = 0,
\eeq
since the boundary conditions do not depend on $F$, and furthermore
\begin{align*}
\frac{\partial \Phi_i}{\partial F_{i-1}}
& = -\frac{1}{6} h_{i-1} + \frac{1}{12} J_{i - 1/2} h_{i-1}^2 \\
\frac{\partial \Phi_i}{\partial F_{i-1/2}}
& = -\frac{2}{3} h_{i-1} \\
\frac{\partial \Phi_i}{\partial F_{i}}
& = -\frac{1}{6} h_{i-1} - \frac{1}{12} J_{i - 1/2} h_{i-1}^2
\end{align*}
for $i = 1 \ldots N$, where
\begin{multline}
\labeln{equ:360}
J_{i - 1/2} \\
= \frac{\partial F}{\partial z}
\left( t_{i-1} + \frac{h_{i-1}}{2}, \frac{z_{i-1} + z_{i}}{2} - \frac{h_{i-1}}{8} \left( F_i - F_{i-1}\right)\right).
\end{multline}
The algorithm \texttt{bvp4c} provides a numerical approximation to this quantity.
These relations as well as the inverse of the collocation Jacobian (also available from \texttt{bvp4c}) can be used to compute the sensitivity through Equation~\eqref{equ:395}.
\end{document}